\documentclass[aps,prd,preprint,draft,showpacs]{revtex4}
\usepackage{graphicx,epsf,amssymb}
\newcommand{\be}{\begin{equation}} 
\newcommand{\ee}{\end{equation}}

\newcommand{\bea}{\begin{eqnarray}} 
\newcommand{\eea}{\end{eqnarray}} 
\newcommand{\Tr}{{\rm Tr}}

\newcommand{\NeqFour}{{\cal N} =4}

\def\adot{{\dot a}}
\def\bdot{{\dot b}}
\def\aldot{{\vphantom{\dot\beta}\dot\alpha}}
\def\bedot{{\dot\beta}}
\def\del{\partial}

\def\CP{\mathbb{CP}}

\newif\ifdraft
\draftfalse
\newif\ifpreprint
\preprinttrue

\def\sect#1{section~{\ref{#1}}}
\def\app#1{appendix~{\ref{#1}}}

\def\fig#1{fig.~{\ref{#1}}}

\def\tree{{\rm tree}}
\def\pol{\varepsilon}
\def\Tr{\, {\rm Tr}}

\def\hf{{{1\over2}}}
\def\Neqfour{{\cal N}=4}

\def\NeqFour{{\cal N}=4}
\def\NeqOne{{\cal N}=1}
\def\Numer{{\cal H}}

\def\spa#1.#2{\left\langle#1\,#2\right\rangle}
\def\spb#1.#2{\left[#1\,#2\right]}
\def\sand#1.#2.#3{%
\left\langle\smash{#1}{\vphantom1}^{-}\right|{#2}%
\left|\smash{#3}{\vphantom1}^{-}\right\rangle}
\def\sandp#1.#2.#3{%
\left\langle\smash{#1}{\vphantom1}^{-}\right|{#2}%
\left|\smash{#3}{\vphantom1}^{+}\right\rangle}
\def\sandpp#1.#2.#3{%
\left\langle\smash{#1}{\vphantom1}^{+}\right|{#2}%
\left|\smash{#3}{\vphantom1}^{+}\right\rangle}
\def\sandpm#1.#2.#3{%
\left\langle\smash{#1}{\vphantom1}^{+}\right|{#2}%
\left|\smash{#3}{\vphantom1}^{-}\right\rangle}
\def\sandmp#1.#2.#3{%
\left\langle\smash{#1}{\vphantom1}^{-}\right|{#2}%
\left|\smash{#3}{\vphantom1}^{+}\right\rangle}
\def\sandmm#1.#2.#3{%
\left\langle\smash{#1}{\vphantom1}^{-}\right|{#2}%
\left|\smash{#3}{\vphantom1}^{-}\right\rangle}
\def\spab#1.#2.#3{\sandmm#1.#2.#3}
\def\spba#1.#2.#3{\sandpp#1.#2.#3}
\def\spaa#1.#2.#3.#4{\sandmp#1.{#2#3}.#4}
\def\spbb#1.#2.#3.#4{\sandpm#1.{#2#3}.#4}

\newbox\charbox
\newbox\slabox
\def\s#1{{      
        \setbox\charbox=\hbox{$#1$}
        \setbox\slabox=\hbox{$/$}
        \dimen\charbox=\ht\slabox
        \advance\dimen\charbox by -\dp\slabox
        \advance\dimen\charbox by -\ht\charbox
        \advance\dimen\charbox by \dp\charbox
        \divide\dimen\charbox by 2
        \raise-\dimen\charbox\hbox to \wd\charbox{\hss/\hss}
        \llap{$#1$}
}}

\def\eqn#1{eq.~(\ref{#1})}

\def\eqns#1#2{eqs.~(\ref{#1}) and~(\ref{#2})}

\def\e{\epsilon}
\def\eps{\epsilon}

\def\Gr{{\rm Gr}}

\def\sign{{\mathop{\rm sign}\nolimits}}
\def\lr{\leftrightarrow}

\def\Li{\mathop{\rm Li}\nolimits}
\def\Split{\mathop{\rm Split}\nolimits}
\def\Soft{\mathop{\rm Split}\nolimits}
\def\oneloop{{1 \mbox{-} \rm loop}}

\def\cg{\hat{c}_\Gamma}
\def\Ord{{\cal O}}

\def\sandp#1.#2.#3{%
\left\langle\smash{#1}{\vphantom1}^{+}\right|{#2}%
\left|\smash{#3}{\vphantom1}^{+}\right\rangle}
\def\ksl{\s{k}}
\def\Ksl{\s{K}}

\def\Fact{{\cal F}}
\def\Soft{{\cal S}}

\newbox\ourfigbox
\def\SizedFigureWithCaption#1#2#3{%
\setbox\ourfigbox
  \hbox{\hss\epsfxsize #1 \epsfbox{#2}\hss}
\hbox to \wd\ourfigbox{\vbox{\noindent\copy\ourfigbox\break
\vskip -6mm      \hbox to \wd\ourfigbox{\hss#3\hss}}}
}

\def\spa#1.#2{\left\langle#1\,#2\right\rangle}
\def\spb#1.#2{\left[#1\,#2\right]}
\def\lor#1.#2{\left(#1\,#2\right)}
\def\sand#1.#2.#3{%
\left\langle\smash{#1}{\vphantom1}^{-}\right|{#2}%
\left|\smash{#3}{\vphantom1}^{-}\right\rangle}
\def\sandpp#1.#2.#3{%
\left\langle\smash{#1}{\vphantom1}^{+}\right|{#2}%
\left|\smash{#3}{\vphantom1}^{+}\right\rangle}
\def\sandpm#1.#2.#3{%
\left\langle\smash{#1}{\vphantom1}^{+}\right|{#2}%
\left|\smash{#3}{\vphantom1}^{-}\right\rangle}
\def\sandmp#1.#2.#3{%
\left\langle\smash{#1}{\vphantom1}^{-}\right|{#2}%
\left|\smash{#3}{\vphantom1}^{+}\right\rangle}

\begin{document}
\hfuzz 10 pt


\ifpreprint
\noindent
UCLA/04/TEP/50
\hfill SLAC--PUB--10905
\hfill Saclay/SPhT--T04/166
\fi

\title{All Next-to-Maximally-Helicity-Violating One-Loop \\
Gluon Amplitudes in $\NeqFour$ Super-Yang--Mills Theory%
\footnote{Research supported in part by the US Department of 
 Energy under contracts DE--FG03--91ER40662 and DE--AC02--76SF00515,
 and by the {\it Direction des Sciences de la Mati\`ere\/}
 of the {\it Commissariat \`a l'Energie Atomique\/} of France.}}

\author{Zvi Bern}
\affiliation{ Department of Physics and Astronomy, UCLA\\
\hbox{Los Angeles, CA 90095--1547, USA}
}

\author{Lance J. Dixon} 
\affiliation{ Stanford Linear Accelerator Center \\ 
              Stanford University\\
             Stanford, CA 94309, USA
}

\author{David A. Kosower} 
\affiliation{Service de Physique Th\'eorique, CEA--Saclay\\ 
          F--91191 Gif-sur-Yvette cedex, France
}

\date{December 17, 2004}

\begin{abstract}
We compute the next-to-MHV one-loop $n$-gluon amplitudes in $\NeqFour$
super-Yang--Mills theory.  These amplitudes contain
three negative-helicity gluons and an arbitrary number of
positive-helicity gluons, and are the first infinite series of
amplitudes beyond the simplest, MHV, amplitudes.  We also discuss some
aspects of their twistor-space structure.
\end{abstract}

\pacs{11.15.Bt, 11.30.Pb, 11.55.Bq, 12.38.Bx \hspace{1cm}}

\maketitle



\renewcommand{\thefootnote}{\arabic{footnote}}
\setcounter{footnote}{0}


\section{Introduction}
\label{IntroSection}

The computation of on-shell amplitudes in the maximally supersymmetric
($\NeqFour$) gauge theory (MSYM) has proven to be a useful laboratory for 
developing
computational techniques in perturbative gauge theories.  Explicit results
for amplitudes have in turn assisted the development of Witten's recent
twistor-space topological string theory~\cite{WittenTopologicalString,RSV}, 
a candidate for a weak--weak
dual to the supersymmetric gauge theory.
This string theory generalizes Nair's earlier
description~\cite{Nair} of the simplest gauge-theory amplitudes.
(Berkovits and Motl~\cite{Berkovits,BerkovitsMotl}, Neitzke and
Vafa~\cite{Vafa}, and Siegel~\cite{Siegel} have given alternative
descriptions of the candidate topological string theory.)

One-loop amplitudes in the maximally-supersymmetric theory can also be
regarded as terms in a computation of amplitudes in perturbative QCD.
In particular, the amplitudes where all external states are gluons can
be decomposed into three terms, corresponding to the amplitude in the
$\NeqFour$ theory; to the contribution of a matter multiplet in the
$\NeqOne$ supersymmetric theory; and to the contribution of a scalar
circulating in the loop. Moreover, in special cases, we show
that coefficients of some integral functions in $\NeqFour$ gauge theory 
are identical to those of QCD.   

At tree level, three infinite sequences of gluon amplitudes were
conjectured by Parke and Taylor~\cite{ParkeTaylor}
 in the mid 1980s, and quickly proven
by Berends and Giele~\cite{Recurrence}.
  Amplitudes with zero or one negative-helicity
gluons, and an arbitrary number of positive helicity, vanish.
Amplitudes with two negative-helicity gluons, so-called `MHV'
amplitudes, have a very simple form. 

Investigations of the twistor-space structure of known analytic
results for more complicated helicity patterns led Cachazo,
Svr\v{c}ek, and Witten (CSW)~\cite{CSW} to formulate a new set of
rules for computing tree amplitudes in gauge theories.  These rules
employ vertices that are particular off-shell continuations of the MHV
amplitudes.  The vertices are sewn together using scalar propagators.
These rules have made it straightforward to write down new analytic
expressions for infinite sequences of amplitudes and currents with
three or more negative-helicity legs, that is helicity configurations
beyond MHV~\cite{CSW,Khoze,NMHVTree,OtherCSW,BBK}.  They have also
been used to obtain amplitudes containing a Higgs boson coupled to QCD
via a massive top-quark loop (in the infinite-mass limit)~\cite{Higgs}
and to obtain electroweak vector boson currents~\cite{Currents}.  A
natural question is whether one can compute similar amplitudes at one
loop, and what light they shed on the structure of the twistor-space
string dual to the gauge theory.

The unitarity-based
method~\cite{Neq4Oneloop,Neq1Oneloop,UnitarityMachinery,TwoLoopSplitting}
makes use of the simple forms of tree amplitudes to produce, in turn,
simple forms for infinite sequences of one-loop amplitudes.  In this
approach, we sew together products of on-shell tree amplitudes, and
directly reconstruct Feynman integrals with the same analyticity properties.
It makes
use of the (standard) cuts of amplitudes, corresponding to the
absorptive parts of amplitudes, and also
introduces the non-standard notion of generalized
cuts~\cite{ZFourPartons,TwoLoopUnitarity,TwoLoopSplitting,Neq47} which has
been used effectively in a variety of one- and two-loop calculations.
We have employed both standard and generalized cuts 
for the calculations described
in this paper.  The unitarity-based techniques are 
enhanced by combining them with knowledge of the basis of
dimensionally regularized one-loop integral functions that can appear
in the results~\cite{Integrals5,IntegralsN,Neq4Oneloop}. 
The basis required for one-loop $\NeqFour$ super-Yang--Mills
amplitudes is reproduced in \app{IntegralsAppendix}.
Knowledge of the basis reduces the problem to one of determining 
the coefficients in front of the integral functions.  We have
also made use of the requirement that the infrared divergences
match the known universal form~\cite{UniversalIR} for parts of
the computation.

Recently, stimulated in part by the computation by Brandhuber, Spence
and Travaglini~\cite{BST} of the $\Neqfour$ MHV amplitudes from CSW
diagrams~\cite{CSW}, there has been a great deal of progress in
obtaining and analyzing one-loop amplitudes in $\NeqFour$ and
$\NeqOne$ theories using the unitarity method and twistor-motivated
ideas~\cite{CSWII,CSWIII,BBKR,Cachazo,
BCF7,BCFcoplanar,Neq47,OtherNeqOne,DunbarNeq1,BCF4m,BBSTNeq0}.  These new
results have also made it clear that the simplicity of tree amplitudes
is inherited by their one-loop counterparts.

In this paper, we shall compute all next-to-MHV one-loop
$n$-gluon amplitudes, that is one-loop amplitudes with
three negative-helicity gluons and $(n-3)$ of positive helicity.
Some of the all-$n$ coefficients appearing in the amplitudes were also 
computed elsewhere~\cite{Cachazo,Neq47,BCF4m}
These amplitudes in $\NeqFour$ super-Yang--Mills
theory were computed previously for $n=6$ in ref.~\cite{Neq1Oneloop},
and for $n=7$ in refs.~\cite{BCF7,Neq47}.
For these two cases, using parity one can reduce the number of negative
helicities to three or less;  hence the next-to-MHV amplitudes
exhaust the set of non-MHV amplitudes.

As a by-product of our computation, we have uncovered new
representations of the NMHV $n$-point tree amplitudes. These
representations arise from the required form of infrared divergences in
any one-loop amplitude~\cite{UniversalIR}.  These new representations
suggest that there is a more general formalism than MHV vertices for
systematically and directly generating the tree amplitudes.  The
equivalence between the different representations appears to require
a stronger symmetry than the gauge invariance needed to remove the CSW
reference momentum.

This paper is organized as follows.  In \sect{NotationSection}, we
describe our notation.  Our calculational approach is discussed in
\sect{CalculationSection}, with the results given in
\sect{ResultsSection}.  Consistency checks on the results, as well as
the derivation of a few sets of coefficients from collinear
limits of ones obtained by direct calculation, are described in
\sect{ConsistencySection}. The new representations of $n$-point tree
amplitudes, obtained from the infrared-divergent terms, are presented
in \sect{TreeSection}.  Finally, we discuss twistor-space properties
of the box coefficients, for NMHV and more general amplitudes, in
\sect{TwistorSection}. Our conclusions and outlook are presented in
\sect{ConclusionSection}. We also include two appendices.  The first
contains an explicit demonstration that our all-$n$ box coefficients
are co-planar, as required~\cite{Neq47,BCFcoplanar}.  The second
contains the basis of box integral functions.


\section{Notation}
\label{NotationSection}

We use the trace-based color decomposition~\cite{TreeColor,TreeReview} 
of amplitudes.  At tree level, this decomposition is,
\begin{equation}
{\cal A}_n^\tree(\{k_i,h_i,a_i\}) = 
\sum_{\sigma \in S_n/Z_n} \Tr(T^{a_{\sigma(1)}}\cdots T^{a_{\sigma(n)}})\,
A_n^\tree(\sigma(1^{h_1},\ldots,n^{h_n}))\,,
\label{TreeColorDecomposition}
\end{equation}
where $S_n/Z_n$ is the group of non-cyclic permutations on $n$
symbols, and $j^{h_j}$ denotes the $j$-th momentum and helicity $h_j$.
The $T^a$ are fundamental representation SU$(N_c)$ color matrices
normalized so that $\Tr(T^a T^b) = \delta^{ab}$.  The color-ordered
amplitude $A_n^\tree$ is invariant under a cyclic permutation of its
arguments.

We describe the amplitudes using the spinor helicity formalism.
In this formalism amplitudes are expressed in terms of spinor
inner-products,
\begin{equation}
\spa{j}.{l} = \langle j^- | l^+ \rangle = \bar{u}_-(k_j) u_+(k_l)\,, 
\hskip 2 cm
\spb{j}.{l} = \langle j^+ | l^- \rangle = \bar{u}_+(k_j) u_-(k_l)\, ,
\label{spinorproddef}
\end{equation}
where $u_\pm(k)$ is a massless Weyl spinor with momentum $k$ and plus
or minus chirality~\cite{SpinorHelicity,TreeReview}. Our convention
is that all legs are outgoing. The notation used here follows the
standard QCD literature, with $\spb{i}.{j} = \sign(k_i^0 k_j^0)\spa{j}.{i}^*$
so that,
\begin{equation}
\spa{i}.{j} \spb{j}.{i} = 2 k_i \cdot k_j = s_{ij}\,.
\end{equation}
(Note that the square bracket $\spb{i}.{j}$ differs by an overall sign
compared to the notation commonly used in twistor-space
studies~\cite{WittenTopologicalString}.) 

\def\vmu{{\vphantom{\mu}}}
We denote the sums of cyclicly-consecutive external momenta by
\begin{equation}
K^\mu_{i\ldots j} \equiv 
   k_i^\mu + k_{i+1}^\mu + \cdots + k_{j-1}^\mu + k_j^\mu,
\label{KDef}
\end{equation}
where all indices are mod $n$ for an $n$-gluon amplitude.
The invariant mass of this vector is $s_{i\ldots j} = K_{i\ldots j}^2$.
Special cases include the two- and three-particle invariant masses, 
which are denoted by
\begin{equation}
s_{ij} \equiv (k_i+k_j)^2 = 2k_i\cdot k_j,
\qquad \quad
s_{ijk} \equiv (k_i+k_j+k_k)^2.
\label{TwoThreeMassInvariants}
\end{equation}
In color-ordered amplitudes only invariants with cyclicly-consecutive
arguments need appear, {\it e.g.}{} $s_{i,i+1}$ and $s_{i,i+1,i+2}$.
We also write, for the sum of massless momenta belonging to a set $A$,
\be
K^\mu_A \equiv \sum_{a_i \in A}
   k_{a_i}^\mu \,.
\label{KDefAlt}
\end{equation}
(The sets that will appear in explicit expressions will be of
cyclicly consecutive external momenta.)
For non-MHV loop amplitudes, longer spinor strings than~(\ref{spinorproddef})
will typically appear, such as
\begin{equation}
 \spba{i}.{\Ksl_A}.{j}  \qquad  \hbox{and} \qquad
 \spaa{i}.{\Ksl_A}.{\Ksl_B}.j
  \,.
\label{longerstrings}
\end{equation}

The simplest color-ordered amplitudes are the maximally
helicity-violating (MHV) Parke-Taylor tree amplitudes~\cite{ParkeTaylor},
which have two negative-helicity gluons and the rest of positive helicity,
\begin{equation}
  A^{\tree \rm\ MHV}_{m_1m_2}(1,2,\ldots,n)
 =  i\, { {\spa{m_1}.{m_2}}^4 \over \spa1.2\spa2.3\cdots\spa{n}.1 }\, ,
\label{PT}
\end{equation}
where $m_{1,2}$ label the negative-helicity legs.

For one-loop amplitudes, the color decomposition is
similar to the tree-level
case~(\ref{TreeColorDecomposition})~\cite{BKColor}.  
When all internal particles transform in
the adjoint representation of SU$(N_c)$, as is the case for 
$\NeqFour$ supersymmetric Yang--Mills theory, we have
\begin{equation}
{\cal A}_n^\oneloop ( \{k_i,h_i,a_i\} ) =
  \sum_{c=1}^{\lfloor{n/2}\rfloor+1}
      \sum_{\sigma \in S_n/S_{n;c}}
     \Gr_{n;c}( \sigma ) \,A_{n;c}(\sigma) \,,
\label{ColorDecomposition}
\end{equation}
where ${\lfloor{x}\rfloor}$ is the largest integer less than or equal to $x$.
The leading color-structure factor
\begin{equation}
\Gr_{n;1}(1) = N_c\ \Tr (T^{a_1}\cdots T^{a_n} ) \,, 
\end{equation}
is $N_c$ times the tree color factor.  The subleading color
structures are given by
\begin{equation}
\Gr_{n;c}(1) = \Tr ( T^{a_1}\cdots T^{a_{c-1}} )\,
\Tr ( T^{a_c}\cdots T^{a_n}).
\end{equation}
$S_n$ is the set of all permutations of $n$ objects,
and $S_{n;c}$ is the subset leaving $\Gr_{n;c}$ invariant.

The one-loop subleading-color partial amplitudes are given by a sum over
permutations of the leading-color ones~\cite{Neq4Oneloop}.
Therefore we need to compute directly only the leading-color 
single-trace partial amplitudes $A_{n;1}$. 

\begin{figure}[t]
\centerline{\epsfxsize 6.0 truein \epsfbox{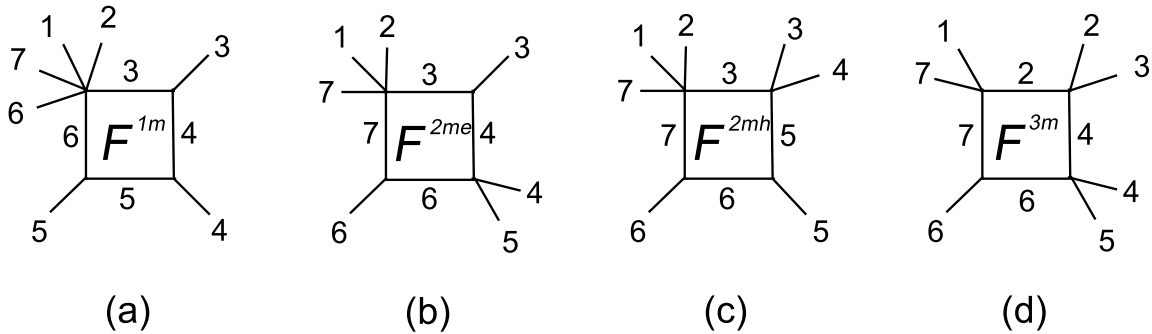}}
\caption[a]{\small Examples of box integral functions $B(i,j,k,l)$
appearing in seven-point amplitudes; the arguments $i,j,k,l$ are the
uncanceled propagators: 
(a) the one-mass box $B(3,4,5,6)=F^{\rm 1m}(s_{34},s_{45},s_{345})$,
(b) the `easy' two-mass box 
$B(3,4,6,7) = F^{{\rm 2m}e}(s_{345},s_{456},s_{45},s_{712})$,
(c) the `hard' two-mass box
$B(3,5,6,7) = F^{{\rm 2m}h}(s_{56},s_{345},s_{712},s_{34})$,
and (d) the three-mass box
$B(2,4,6,7) = F^{{\rm 3m}}(s_{671},s_{456},s_{71},s_{23},s_{45})$.
}
\label{boxexamplesFigure}
\end{figure}

The $\NeqFour$ SYM amplitudes may be expressed as a sum of scalar box
integrals ${\cal I}_4$, multiplied by coefficients which are rational
functions of spinor products~\cite{Neq4Oneloop}.  It is convenient to
multiply these integrals by suitable dimensionful combinations of
kinematic invariants in order to obtain `box functions' $F$ whose
series expansions in $\e$ only contain logarithmic or polylogarithmic
dependence on the kinematic invariants.  The necessary box functions
$F^{4{\rm m}}$, $F^{3{\rm m}}$, $F^{2{\rm m} \, h}$, $F^{2{\rm m}\, e}$
and $F^{1{\rm m}}$ (and for $n=4$, $F^{0{\rm m}}$), are listed in
\app{IntegralsAppendix}.
  The kinematics of each
box function appearing in an $n$-point amplitude is determined by
canceling $(n-4)$ propagators from the $n$-point diagram with
external legs in the order 1,2,3,$\ldots,n$.  In ref.~\cite{Neq47} we
labeled the box integrals for $n=7$ by a triplet of integers, say
$(i',j',k')$, corresponding to the three propagators canceled from
the heptagon diagram with external legs in the order $1,2,3,\ldots,7$.
This labeling scheme becomes very cumbersome for discussing the
all-$n$ case, since the number of integers required grows with $n$.
Here we therefore choose to label the integrals, and their kinematic
coefficients, by a {\it quartet} $(i,j,k,l)$ of distinct integers,
corresponding to the four {\it un}canceled propagators.  In the
seven-point case, this quartet is the complement of the 
triplet~$(i',j',k')$ used
in ref.~\cite{Neq47}, $\{i,j,k,l\} \cup \{i',j',k'\} =
\{1,2,3,4,5,6,7\}$.  See \fig{boxexamplesFigure} for examples of this
labeling.   (We also use the notation $B(i,j,k,l)$ for the labeled
box functions, instead of $F(i,j,k,l)$, in order to avoid confusion
with the twistor-space co-linear operator $F_{ijk}$ discussed in 
\sect{TwistorOverviewSubsection}.)

We write the $\NeqFour$ leading-color partial amplitude
as~\cite{Integrals5,IntegralsN,Neq4Oneloop}
\be
A_{n;1}^{\Neqfour} = i \cg \, (\mu^2)^\e
\sum_{i,j,k,l} c_{ijkl} B(i,j,k,l)\,,
\label{GenBoxDecomp}
\ee
where $c_{ijkl}$ is the kinematic coefficient and
\be
\cg\ =\ {1 \over (4 \pi)^{2-\e}}
{\Gamma(1+\e)\Gamma^2(1-\e)\over\Gamma(1-2\e)}
\label{cGamma}
\ee
is a ubiquitous prefactor, and $\mu$ is the trivial scale dependence
of all dimensionally-regulated
one-loop amplitudes.  While the $\NeqFour$ theory is ultraviolet-finite,
on-shell amplitudes still have infrared divergences which are also regulated
dimensionally, and sneak in a dependence on $\mu$.

For a given helicity amplitude, the number of box functions, and box
coefficients, is the number of un-ordered quartets of distinct integers
$(i,j,k,l)$ with each integer running from 1 to $n$, and all four unequal.
This number is just $( {n \atop 4} )$.  These include,
\begin{itemize}
\item one-mass boxes shown in \fig{boxexamplesFigure}a ($n$ boxes),
\item the easy two-mass boxes shown in \fig{boxexamplesFigure}b, plus
cyclic permutations ($n (n-5)/2$ boxes in total),
\item the hard two-mass boxes shown in \fig{boxexamplesFigure}c 
($n (n-5)$ boxes),
\item the three-mass box shown in \fig{boxexamplesFigure}d 
($n (n-5) (n-6)/2$ boxes).
\end{itemize}

We take the three negative helicity gluons to be labeled by 
$m_1,m_2,m_3$.


\section{Calculational Approach}
\label{CalculationSection}

Computing an infinite series of amplitudes would require computing an
infinite number of Feynman diagrams.  The unitarity-based method,
 however, can reduce such a computation to a finite
one.  We use it.

\begin{figure}[t]
\centerline{\epsfxsize 2.25 truein \epsfbox{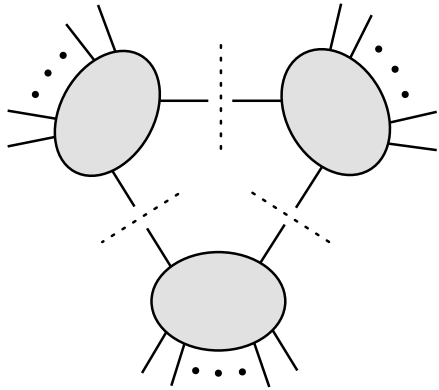}}
\caption[a]{\small A generalized triple cut.  The three
propagators cut by the dashed lines are required to be `open'.
}
\label{triplecutFigure}
\end{figure}

In the unitarity-based method, we reconstruct a loop amplitude from
tree amplitudes by requiring that internal propagators go on shell.
Letting two propagators go on shell corresponds to extracting
the absorptive parts, which are just phase-space integrals over products
of tree amplitudes.  In most cases it is convenient to also use generalized
cuts~\cite{ZFourPartons,TwoLoopUnitarity,TwoLoopSplitting,Neq47}, where
multiple propagators go on shell, 
such as the triple cut shown in~\fig{triplecutFigure}. 
Taking a generalized cut corresponds
to extracting those contributions to a loop amplitude where all cut
propagators are required to be present.
 The generalized cuts have the
property of reducing the building blocks of loop computations to the
simplest possible set of tree amplitudes.  They can even reduce 
higher-loop
calculations to integrals over products of tree
amplitudes~\cite{TwoLoopUnitarity,TwoLoopSplitting}.  In all cases,
one reconstructs the loop integrals giving rise to the required
ordinary or generalized cuts.  In the present calculation, that 
only requires identifying the appropriate integral in the basis set.

Our coefficients were entirely obtained from the generalized triple
and quadruple cuts by augmenting them with infrared consistency
conditions as well as collinear and soft behaviors. The soft and
collinear limits allow us to obtain unknown coefficients from
explicitly computed ones.  
When required infinite series of tree-level amplitudes
are known, the unitarity method enables us to compute infinite series
of one-loop amplitudes.
The combination of the various methods allows us to
give explicit formul\ae{} for all coefficients in NMHV $n$-point
amplitudes.

In general, we must compute these cuts in $D$
dimensions.  For one-loop amplitudes 
in supersymmetric theories, however, it suffices to
compute them in four dimensions~\cite{Neq4Oneloop,Neq1Oneloop}.  
(The reconstructed loop integral is still computed in $D$ dimensions, 
of course.)
The four-dimensional amplitudes are most efficiently and conveniently
handled in a helicity basis.  Starting from tree {\it amplitudes\/} rather
than {\it diagrams\/} means that the extensive cancellations that occur in
gauge theories are taken into account before any loop integrations are
done, which greatly reduces the complexity of the calculations.

In extracting the triple cut, terms which vanish as the cut propagators
go on shell may be dropped.  The utility of this procedure comes
from two aspects: the triple cut itself may be represented as a product
of three tree amplitudes; and the resulting expression isolates coefficients
of a more limited class of integrals than the ordinary (absorptive) cut. 
 In the calculations
we perform, these coefficients turn out to be the simplest of all integral
coefficients (and even simple in an absolute sense).  All three-mass
and hard two-mass box coefficients may be determined from a triple cut.
In an NMHV loop amplitude, each tree amplitude making up a triple cut
will be an MHV amplitude, that is with two negative-helicity gluons, be
they external or (cut) internal ones.  

The quadruple cuts show very simply that all four-mass box
coefficients must vanish in an NMHV amplitude. These cuts are given
by products of four tree amplitudes.  However, there are only 
seven negative-helicity gluon legs available:
three are external gluons, and four are gluons crossing the cuts 
(one for each of the four cuts).  Hence at least one of the four
tree amplitudes must have fewer than two negative-helicity gluons, and
will therefore vanish.

Having determined the three-mass and hard two-mass box coefficients
from the triple cuts (or equivalently the ordinary cuts), we can
determine the easy two-mass and one-mass coefficients in two
independent ways.  The first is to return to the `ordinary' cuts, and
compute them.  In this case, we will have the product of an MHV and an
NMHV amplitude forming the cut.  Depending on the configuration of the
external negative-helicity gluons (and on the channel we cut),
contributions will come either from gluons alone circulating in the
loop (a `singlet' contribution) or from all states in the $\NeqFour$
multiplet (`non-singlet').  The tree-level CSW rules make it easy to
write down analytic expressions for the NMHV amplitudes, but
unfortunately the form they yield --- containing off-shell momenta
either in the original CSW form or in the modified `projected' form
--- is not directly suitable for use in the unitarity-based method,
because it is not clear what propagators should be reconstructed from
these unusual denominators.  For the gluon amplitudes, however, a
corresponding expression in terms of on-shell spinor products and
invariants alone is known~\cite{NMHVTree}, and it is this expression
we have used in computing the cuts.  This provides a computation of
the (subset of) easy two-mass and one-mass box coefficients that have
singlet cuts, that is which have a cut that isolates all three
negative helicities on one side of the cut. 
The calculation starts with an octagon integrand,
which reduces to a sum of box integrands via spinor algebra and the
introduction of appropriate `cubic' invariants~\cite{Neq47}.
Brute-force integral reduction techniques (e.g.{} Brown--Feynman or
Passarino--Veltman~\cite{BFPV}), which introduce nasty spurious
Gram-determinant denominators, were not required.

The other method of determining these coefficients is to use the
infrared consistency equations.  These equations arise from
confronting our knowledge of the structure of infrared singularities
in the amplitude with the presence of singularities in
individual box functions.  On general grounds~\cite{UniversalIR}, we
know that only nearest-neighbor two-particle invariants can appear in
infrared-singular terms, which have the form,
\begin{equation}
A_{n;1}^{\NeqFour} \Bigr|_{\e\ {\rm pole}} =  
-{\cg\over\e^2} \sum_{i=1}^n \biggr({\mu^2\over -s_{i,i+1}}\biggl)^{\e}
\times A_n^{\tree} \,,
\label{OneLoopIRPoles}
\end{equation}
where $\mu$ is an arbitrary scale.  On the other hand, the box
functions listed in \app{IntegralsAppendix} contain singularities
with coefficients $s^{-\e}$ for a much larger set of invariants $s$.
In general, \eqn{OneLoopIRPoles} implies that the coefficient of any
given $\ln(-s_{i,i+1})/\e$ must be equal to the tree; and the
coefficient of any other $\ln(-s_{i\ldots j})/\e$ must vanish.  Both
types of equation are non-trivial.  There are $n(n-3)/2$
such equations corresponding to the number of independent kinematic
invariants. Each box function, listed in \app{IntegralsAppendix},
contains various $\ln(-s_{i,i+1})/\e$ and $\ln(-s_{i,i+1,i+2})/\e$
terms with coefficients 0, $\pm1$ or $\pm{1\over2}$.  The constraints
arising from~\eqn{OneLoopIRPoles} thus become simple linear relations
among the coefficients, some of which involve the tree amplitude.  As
mentioned at the end of \sect{NotationSection}, 
there are a total of $n$ one-mass
boxes and $n (n-5)/2$ easy two-mass boxes, which together precisely
match the number of infrared consistency equations.

It turns out that for $n$ odd the system of equations is
non-degenerate (verified numerically up to $n = 29$), 
so using the infrared consistency
equations we can solve for {\it all} easy two-mass and one-mass box
coefficients in terms of the three-mass and hard two-mass ones.
  For $n$ even it turns out that there is one redundant
equation, so that we can solve for all but one easy two-mass
or one-mass box coefficient.  Of course, once we have obtained
the solution for odd 
$n$, we can confirm that the solution also holds for even $n$ by taking
collinear limits, as we shall mention
in \sect{ConsistencySection}.  The solutions
obtained from the infrared consistency equations, as it turns out,
yield a simpler analytic (but numerically identical) form for the
singlet coefficients than the direct computation discussed above.  We
have also used these infrared consistency equations to obtain the
non-singlet easy two-mass and one-mass coefficients.


\section{Results}
\label{ResultsSection}

In this section we present the results for the box coefficients $c_{ijkl}$
appearing in \eqn{GenBoxDecomp}.  It is convenient to label the coefficients
in terms of clusters. For $X=A,B,C$, let $X_1$ denote the 
first massless momentum in $X$, and $X_{-1}$ the last massless momentum.

\begin{figure}[t]
\centerline{\epsfxsize 6. truein \epsfbox{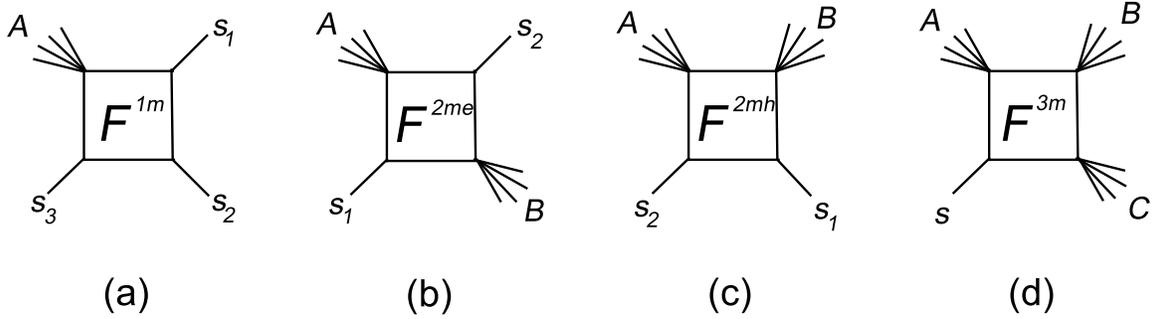}}
\caption[a]{\small The box integral functions labeled by the
clusters of masses.}
\label{clusterlabelsFigure}
\end{figure}

As mentioned in the previous section, from the generalized quadruple 
cuts we see that the four-mass box coefficients all vanish.

The three-mass box coefficients are all given by a single `term'.
This simplicity is tied to the very constrained twistor-space
structure of such coefficients.  A three-mass box integral has a
unique massless `singlet' leg $s$, followed clockwise
around the loop by three massive clusters $A$, $B$ and $C$, as shown
in \fig{clusterlabelsFigure}d.  (The
momenta within each cluster are of course also ordered clockwise.)
Then the three-mass box coefficient is given by,
\bea
c^{3{\rm m}}(m_1,m_2,m_3;s,A,B,C) 
&=& 
{ \bigl[ \Numer(m_1,m_2,m_3;s,A,B,C) \bigr]^4
   \over \spa1.2\spa2.3\cdots\spa{n}.1 \ K_B^2 }
\nonumber \\ \hskip0.0cm
&& \times
{ \spa{A_{-1}}.{B_1} \over
  \sandmp{s}.{\Ksl_C\Ksl_B}.{A_{-1}}
  \sandmp{s}.{\Ksl_C\Ksl_B}.{B_1} } 
\nonumber \\ \hskip0.0cm
&& \times
{ \spa{B_{-1}}.{C_1} \over
  \sandmp{s}.{\Ksl_A\Ksl_B}.{B_{-1}}
  \sandmp{s}.{\Ksl_A\Ksl_B}.{C_1} } \,,
\label{threemassc}
\eea
where all the dependence on the $m_i$ is contained within $\Numer$.
In the cases where the singlet leg $s$ has positive helicity, we find,
\bea
\Numer
&=& 0, \hskip6.5cm \hbox{$m_{1,2,3} \in A$,}
\label{NAAA} \\
&=& 0, \hskip6.5cm \hbox{$m_{1,2,3} \in B$,}
\label{NBBB} \\
&=& \spa{m_1}.{m_2} \sandmp{s}.{\Ksl_C\Ksl_B}.{m_3}, 
\hskip2.1cm \hbox{$m_{1,2} \in A$, $m_3 \in B$,}
\label{NAAB} \\
&=& \spa{m_1}.{m_2} \spa{s}.{m_3} K_B^2, 
\hskip3.5cm \hbox{$m_{1,2} \in A$, $m_3 \in C$,}
\label{NAAC} \\
&=& \spa{m_1}.{m_2} \sandmp{s}.{\Ksl_C\Ksl_B}.{m_3}, 
\hskip2.1cm \hbox{$m_{1,2} \in B$, $m_3 \in A$,}
\label{NABB} \\
&=&    \spa{m_1}.{m_2} \sandmp{s}.{\Ksl_A \Ksl_B}.{m_3}
\nonumber \\ && \hskip0.8cm
        + \spa{m_3}.{m_2} \sandmp{s}.{\Ksl_C \Ksl_B}.{m_1},
\hskip1.0cm \hbox{$m_1 \in A$, $m_2 \in B$, $m_3 \in C$,}
\label{NABC}
\eea
plus cases obtained by exchanging $A$ and $C$ (using reflection/flip
symmetry).
An alternative form for the last case is,
\bea
\Numer
&=&    \spa{s}.{m_1} \sandmp{m_3}.{(\ksl_s+\Ksl_C)\Ksl_B}.{m_2}
\nonumber \\ && \hskip0.8cm
     + \spa{s}.{m_3} \sandmp{m_1}.{\Ksl_A\Ksl_B}.{m_2}, 
\hskip1.0cm \hbox{$m_1 \in A$, $m_2 \in B$, $m_3 \in C$.}
\label{NABCAlt}
\eea
In the cases where the singlet leg has negative helicity, $s=m_3$, 
we find,
\bea
\Numer
&=& 0, \hskip7.2cm \hbox{$m_{1,2} \in A$,}
\label{NsAA} \\
&=& \spa{m_1}.{m_2} \sandmp{s}.{\Ksl_C\Ksl_B}.{s}, 
\hskip3.15cm \hbox{$m_{1,2} \in B$,}
\label{NsBB} \\
&=& \spa{s}.{m_1} \sandmp{s}.{\Ksl_C\Ksl_B}.{m_2}, 
\hskip3.2cm \hbox{$m_1 \in A$, $m_2 \in B$,}
\label{NsAB} \\
&=& \spa{s}.{m_1} \spa{s}.{m_2} \, K_B^2,
\hskip4.55cm \hbox{$m_1 \in A$, $m_2 \in C$,}
\label{NsAC}
\eea
plus cases obtained by exchanging $A$ and $C$.

All the other box coefficients are given by appropriate sums of
$c^{3{\rm m}}$ quantities.  In many instances, \eqn{threemassc} will
then be required when the set $A$ or $C$ `degenerates' to a single leg.
($X_{1} = X_{-1} = X$ if the cluster degenerates to a single massless momentum.)
The formula is perfectly well-defined in this limit.
On the other hand, the set $B$ will never be allowed to degenerate 
to a single leg, because the $K_B^2$ factor in the denominator of 
\eqn{threemassc} would then vanish.  In the following equations,
$m_{1,2,3}$ do not play any distinguished role, and for simplicity
we shall suppress these arguments.

The hard two-mass boxes are defined by two adjacent singlet legs,
$s_1$ and $s_2$, followed by two adjacent massive clusters $A$ and $B$,
as shown in \fig{clusterlabelsFigure}c.
Their coefficients are given simply by the sum of two $c^{3{\rm m}}$s:
\be
c^{{\rm 2m}h}(s_1,s_2,A,B) 
= c^{3{\rm m}}(s_1,\{s_2\},A,B)
+ c^{3{\rm m}}(s_2,A,B,\{s_1\}) \,.
\label{twomasshc}
\ee
In \sect{ConsistencySection} we will confirm this formula using
soft limits. 

The easy two-mass boxes are defined by a singlet leg $s_1$ 
followed cyclicly by a massive cluster $A$, then another singlet leg
$s_2$, then a final massive cluster $B$, 
as shown in \fig{clusterlabelsFigure}b.
They are given
by a pair of double sums over $c^{3{\rm m}}$ coefficients.
In the first sum, leg $s_1$ is treated as a singlet, the
first massive cluster must include $s_2$, and the second massive 
cluster must not degenerate to a massless leg.  
Otherwise there are no restrictions
on the sum.  The second sum can be obtained from the first sum by
exchanging the roles of $s_1 \lr s_2$ and $A \lr B$.  The result is,
\bea
c^{{\rm 2m}e}(s_1,A,s_2,B) 
&=& \sum_{k=0}^{M(s_1,s_2)} \sum_{l=0}^{M(s_1,s_2)-k}
      c^{3{\rm m}}(s_1,\hat{A}(s_1,s_2,k),\hat{B}(s_1,s_2,k,l),\hat{C}(s_1,s_2,k,l))
\nonumber \\ && \hskip-0.5cm 
+ \sum_{k=0}^{M(s_2,s_1)} \sum_{l=0}^{M(s_2,s_1)-k}
      c^{3{\rm m}}(s_2,\hat{A}(s_2,s_1,k),\hat{B}(s_2,s_1,k,l),\hat{C}(s_2,s_1,k,l))
  \,,
\nonumber \\
&& {~}
\label{twomassec}
\eea
where 
\bea
\hat{A}(s_1,s_2,k) &=& \{s_1+1,\ldots,s_2+k\} \,,
\label{Ahat} \\
\hat{B}(s_1,s_2,k,l) &=&  \{s_2+k+1,\ldots,s_2+k+l+2\} \,,
\label{Bhat} \\
\hat{C}(s_1,s_2,k,l) &=&  \{s_2+k+l+3,\ldots,s_1-1\} \,,
\label{Chat}
\eea
and
\be
M(s_1,s_2) = n-4-[(s_2-s_1)\ \hbox{mod $n$}]\,.
\label{twomassesumlimit}
\ee

\begin{figure}[t]
\centerline{\epsfxsize 5. truein \epsfbox{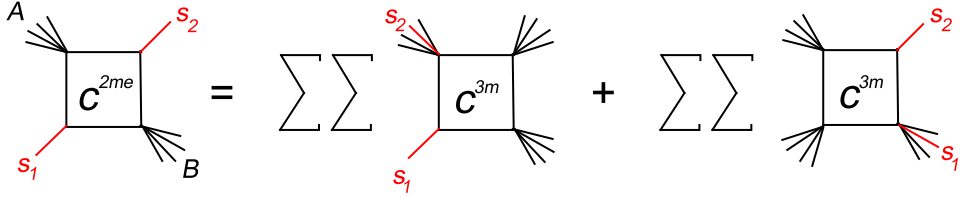}}
\caption[a]{\small Schematic depiction of the easy two-mass box
  coefficients, expressed as a double sum over three-mass
  box coefficients.}
\label{c2meFigure}
\end{figure}

A schematic depiction of the double sum~(\ref{twomassec}) is
provided in \fig{c2meFigure}.  Note that there is a certain
cyclic `handedness' to the sum, in that the `buried' leg $s_2$
is clockwise from the singlet leg $s_1$ in the first sum,
and similarly in the second sum. 
There is an alternative representation where this handedness is reversed,
which we have numerically verified to be equivalent.

The one-mass boxes are defined by three adjacent singlet legs,
$s_1$, $s_2$ and $s_3$, followed by a massive cluster $A$,
as shown in \fig{clusterlabelsFigure}a.
Their coefficients are given by the degeneration of the easy two-mass
formula, plus a single additional term:
\be 
c^{\rm 1m}(s_1,s_2,s_3,A) 
= c^{{\rm 2m}e}(s_1,\{s_2\},s_3,A)
+ c^{3{\rm m}}(s_2,\{s_3\},A,\{s_1\}) \,.
\label{onemassc}
\ee

Many of the above coefficients also carry over to the corresponding
amplitudes in QCD.  We can apply the generalized cuts to determine
which coefficients in QCD are identical to those given above.
Many types of integral functions beyond those appearing
in the $\NeqFour$ results also contribute
to the full QCD results --- scalar triangle integrals, scalar and 
tensor bubble integrals --- and their coefficients are of course
undetermined by the calculations in this paper.  
Let us assume that we are working in a basis of integrals where
the only box integrals appearing are those in $D=4-2\e$.
Whenever a box coefficient 
can be determined from a (generalized) cut in which {\it only gluons\/}
 are allowed to propagate around the loop, then the fermion and scalar
contributions are absent, and the QCD coefficient is exactly the 
same as the coefficient in $\NeqFour$ super-Yang--Mills theory.  
Suppose, for example, a box integral has a massive 
leg out of which only positive-helicity gluons flow.  Inspecting
the cut which separates that leg from the rest of the amplitude,
we see that only gluons can contribute.  
Thus having an `all-plus mass' is a sufficient condition for a box
coefficient in QCD (or in the pure-glue theory) to be determined by 
the $\NeqFour$ formul\ae{} given in this section.
It is worth noting that amongst the three-mass boxes, the only case which 
does {\it not} satisfy this condition is 
$m_1 \in A$, $m_2 \in B$, $m_3 \in C$ (\eqn{NABC}).  
In the cases of the hard two-mass and one-mass boxes, for QCD and
$\NeqFour$ super-Yang-Mills theory
to give the same result, another sufficient 
condition is that two adjacent massless legs have the same helicity.


\section{Consistency of the Results}
\label{ConsistencySection}

We have performed a number of non-trivial checks on the amplitudes.
One simple check is against all previously
computed~\cite{Neq4Oneloop,BCF7,Neq47} $\NeqFour$ amplitudes, when the
number of legs $n$ is taken to be six or seven.  Another check, 
discussed in \sect{CalculationSection} and valid
beyond $n=7$, is our
computation of many of the easy two-mass and one-mass box coefficients
in two independent ways, using the infrared consistency conditions and
also direct computation.

Amplitudes are constrained by a variety of non-trivial requirements.
Their analytic properties are tightly constrained because all
kinematic poles and cuts must correspond to propagation of physical particles.
They must also have infrared singularities corresponding to
the universal emission of soft and collinear gluons.  

In the collinear region, $k_a \to z k_P$, $k_b \to (1-z) k_P$, 
where $k_P$ is the momentum of the quasi-on-shell intermediate state $P$,
with helicity $h$.  In this limit, massless color-ordered tree 
amplitudes behave as
\begin{equation}
A_{n}^{\tree}\ \mathop{\longrightarrow}^{a \parallel b}\
\sum_{h=\pm} 
\Split^\tree_{-h}   (z, a^{h_a},b^{h_b})\,
         A_{n-1}^{\tree}(\ldots(a+b)^h\ldots)\,,
\label{TreeSplit}
\end{equation}
where $\Split^\tree_{-h}$ are tree-level splitting
amplitudes~\cite{TreeReview}. At one loop, the generalization is,
\begin{eqnarray}
&& A_{n;1}^{\oneloop}\ \mathop{\longrightarrow}^{a \parallel b}\
\sum_{h=\pm}  \biggl(
\Split^\tree_{-h}   (z, a^{h_a},b^{h_b})\,
         A_{n-1;1}^{\oneloop}(\ldots(a+b)^h\ldots) \nonumber \\
& & \hskip 2.7 cm \null
 + \Split^{\oneloop}_{-h}(z,a^{h_a},b^{h_b})\,
         A_{n-1}^\tree(\ldots(a+b)^h\ldots) \biggr) \,,
\label{LoopSplit}
\end{eqnarray}
where the $\Split^{\oneloop}_{-h}$ are one-loop splitting
amplitudes, which are tabulated in the second appendix of
ref.~\cite{Neq4Oneloop}.  This reference also contains a
discussion of the behavior of the collinear limits of one-loop
amplitudes and integral functions.   We will refer to the original 
amplitude as the `parent' and the resulting amplitude appearing
in the collinear limit as the `daughter' amplitude.

\begin{figure}[t]
\centerline{\epsfxsize 2. truein \epsfbox{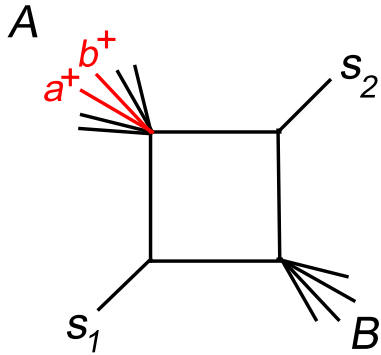}}
\caption[a]{\small An easy two-mass box where positive-helicity 
legs $a$ and $b$ are buried inside a cluster.}
\label{PburiedFigure}
\end{figure}

Besides providing non-trivial checks, 
collinear limits also allow us to fill a small gap in our 
calculation of coefficients using the infrared consistency conditions,
which appears when the number of legs is even.
Recall that as discussed in \sect{CalculationSection}, for even $n$ 
there is one redundant equation, and accordingly we are missing one equation
needed to completely determine 
 the easy two-mass and one-mass box coefficients
(which we collectively refer to as `easy-class').
For odd $n$
we have exactly the right number of equations.  One simple way around
this problem is to prove the correctness of coefficients for even $n$ by
taking the collinear limits of the $(n+1)$-point (odd) case.
Because we are missing only one equation, the
confirmation of even a single easy-class coefficient is sufficient to prove
that our solution is complete.  We can therefore choose the simplest collinear
limits to evaluate.  A suitably simple limit arises 
when the two color-adjacent legs becoming collinear, $a$ and $b$, 
both have positive helicity, and are buried inside a cluster in the 
parent easy two-mass box coefficient, as shown in
\fig{PburiedFigure}.  If $a$ or $b$ is adjacent to one of the
massless legs of the easy two-mass box, the analysis is more
complicated, but we do not need to consider such cases.

According to~\eqn{LoopSplit}, there are contributions proportional to the 
one-loop
splitting function $\Split^{\oneloop}$ as well as those proportional to the
tree splitting function~$\Split^{\tree}$.  
Let us examine the latter terms,
 because their contributions are entirely dictated by
the collinear behavior of the box coefficients $c^{{\rm 2m}e}$.

To understand the collinear limits we need to inspect the
easy two-mass box coefficients.  These coefficients are sums of 
three-mass box coefficients, as given in \eqn{twomassec}.  In a given term in
the sum, if $a$ and $b$ belong to a single mass of the coefficient
$c^{\rm 3m}$ with three or more legs in the mass, it maps very simply
into the terms in the daughter sum because only sums of momenta in the
parent cluster appear in the formula~(\ref{threemassc}), 
{\it i.e.}, $k_a + k_b \rightarrow k_P$
in the daughter coefficient.  We obtain an overall 
tree splitting amplitude factor of $1/(\sqrt{z(1-z)}\spa{a}.{b})$,
coming from the
\begin{equation}
{1\over \spa1.2 \spa2.3 \cdots \spa{n}.1}
\end{equation}
prefactor of the coefficient.  

There are, however, some special cases to consider.  Suppose that in
the given term under consideration in the sum, $a$ and
$b$ are the only legs in one of the massive cluster arguments to $c^{{\rm 3m}}$.
  If the $(a,b)$ cluster
is $A$ or $C$, corresponding to \fig{clusterlabelsFigure}, 
then the limit works as
above, because massless legs are allowed in the sum over three-mass
coefficients if they are in the $A$ or $C$ position.  If $a$ and $b$ are the
only members of $B$, then in the collinear limit there is no
corresponding daughter coefficient in the sum, so all such
coefficients must be non-singular in the collinear limit.  An
investigation of the numerator factors, using eqs.~(\ref{NAAA}),
(\ref{NAAC}), (\ref{NsAA}) and~(\ref{NsAC}), shows that they are indeed
 non-singular and therefore do not contribute.  Finally, for the
term in the three-mass sum where $a$ and $b$ straddle two adjacent
massive legs, either $A$ and $B$, or $B$ and $C$, 
there is also no collinear singularity due to the
factors of $\spa{A_{-1}}.{B_1}$ and $\spa{B_{-1}}.{C_1}$
in \eqn{threemassc}.

There are also contributions to the loop splitting functions.  These
terms arise from discontinuities in the integrals, and also from
hard two-mass and one-mass box integrals~\cite{BernChalmers}.  The
latter terms are easily identifiable because the box integral function 
does not reduce to a daughter integral function (instead it reduces to a
contribution to the loop splitting function).  The issue of
contributions proportional to the loop splitting function is separate
from the contributions proportional to the tree splitting function, so
it is not directly relevant to determining the lower-point
coefficients.  (It could of course be used as an additional check on the
amplitudes.)

The soft limit, in which the momentum of one gluon is scaled to zero, 
provides another check on our results.  (The soft limit may be phrased in
a Lorentz-invariant way as the simultaneous limit $s_{as}, s_{sb}\rightarrow 0$
when $(a,s,b)$ are a sequential triplet of external momenta.)
This limit 
also provides an alternative way to obtain the relation
(\ref{twomasshc}) between the hard two-mass and three-mass
coefficients.  The soft limit of an amplitude obeys an equation
very similar to that for a collinear limit.  At one loop, 
as $k_s \rightarrow 0$,
\begin{eqnarray}
&& A_{n;1}^{\oneloop}(\ldots, s-1, s, s+1, \ldots)\ 
\mathop{\longrightarrow}^{k_s \rightarrow 0}\
\Soft^\tree (s-1, s^{h_s},s+1)\,
         A_{n-1;1}^{\oneloop}(\ldots, s-1, s+1, \ldots) \nonumber \\
& & \hskip 4.7 cm \null
 + \Soft^{\oneloop}(s-1,s^{h_s},s+1)\,
         A_{n-1}^\tree( \ldots, s-1, s+1, \ldots) \,,
\label{LoopSoft}
\end{eqnarray}
where $\Soft^{\tree}$ and $\Soft^{\oneloop}$ are tree and one-loop
soft functions. The tree soft functions are just eikonal
factors~\cite{TreeReview},
\begin{equation}
\Soft^\tree(a,s^+,b) =  {\spa{a}.{b} \over \spa{a}.{s} \spa{s}.b}\,, 
\hskip 2 cm 
\Soft^\tree(a,s^-,b) =  -{\spb{a}.{b} \over \spb{a}.{s} \spb{s}.b}\,.
\end{equation}

As was the case for the collinear limits, 
the contributions proportional to the loop
soft functions are easily separated from the ones
proportional to the tree soft function.  In the $\NeqFour$ theory,
the loop soft functions arise entirely from discontinuities in the box
functions, or from box functions that do not reduce properly 
(map smoothly) to daughter integral functions.  
(Discontinuities arise because of infrared 
divergences~\cite{BernChalmers}).  To obtain
relations between $(n+1)$- and $n$-point coefficients using soft
limits, we need not consider the loop soft functions.  (Again, they can
be used to provide additional consistency checks.)

\begin{figure}[t]
\centerline{\epsfxsize 3. truein \epsfbox{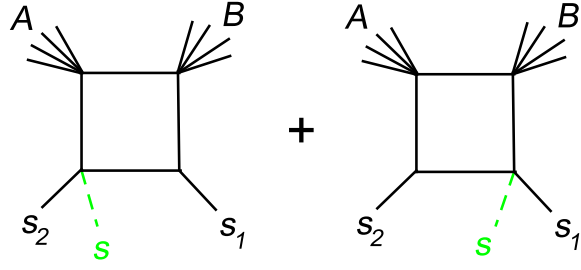}}
\caption[a]{\small The three-mass box integrals that may be used
to verify all hard two-mass coefficients. Leg $s$ becomes soft.}
\label{SoftLimitFigure}
\end{figure}

Consider now the coefficients of the two three-mass boxes displayed in
\fig{SoftLimitFigure}.  As the positive-helicity
leg $s$ becomes soft, it is precisely these
two box functions that reduce to the hard two-mass box function
obtained by simply eliminating leg $s$.  Matching the coefficient of this box
function in the soft limit~(\ref{LoopSoft}) gives the constraint,
\begin{equation}
c^{\rm 3m} (s_1, \{s,s_2\}, A, B) + c^{\rm 3m} (s_2, A, B, \{s_1,s\}) 
\ \mathop{\longrightarrow}^{k_s \rightarrow 0} \
\Soft(s_1, s, s_2) c^{{\rm 2m} h} (s_1, s_2, A, B)\,.
\label{ThreeMassSoftA}
\end{equation}
On the other hand, an inspection of our solution of the three mass 
coefficients reveals that 
\begin{eqnarray}
&& c^{\rm 3m} (s_1, \{s,s_2\}, A, B) + c^{\rm 3m} (s_2, A, B, \{s_1,s\})
 \nonumber \\
&& \hskip 3 cm \null
\ {\mathop{\longrightarrow}^{k_s \rightarrow 0} } \
\Soft(s_1, s, s_2) 
\times \Bigl(
c^{\rm 3m} (s_1, \{s_2\}, A, B) 
 + c^{\rm 3m} (s_2, A,B, \{s_1\}) \Bigr) \,. \hskip 1.5 cm 
\label{ThreeMassSoftB}
\end{eqnarray}
Comparing \eqns{ThreeMassSoftA}{ThreeMassSoftB} then confirms \eqn{twomasshc}
for the hard two-mass box in terms of three-mass coefficients.

The behavior of the NMHV amplitudes under multi-particle factorization 
has an intricate structure which also is useful as a check.  
Here we do not perform a full analysis,
but merely indicate some salient properties.
Let $K^\mu$ denote the cyclicly-adjacent sum of $r$ momenta given by
$K^\mu = (k_i + k_{i+1} + \cdots + k_{i+r-1})^\mu$.
The factorization properties for one-loop amplitudes in the limit 
$K^2 \rightarrow 0$ are described by~\cite{BernChalmers},
\begin{eqnarray}
A_{n;1}^{\oneloop}\
&& \hskip -.42 cm 
 \mathop{\longrightarrow}^{K^2 \rightarrow 0} 
\hskip .15 cm 
\sum_{h=\pm}  \Biggl[
   A_{r+1;1}^{\oneloop}(k_i, \ldots, k_{i+r-1}, K^h) \, {i \over K^2} \, 
   A_{n-r+1}^{\tree}((-K)^{-h}, k_{i+r}, \ldots, k_{i-1})
\nonumber \\
&& \hskip-.6cm \null
 +  A_{r+1}^{\tree}(k_i, \ldots, k_{i+r-1}, K^h) \, {i\over K^2} \,
      A_{n-r+1;1}^{\oneloop}((-K)^{-h}, k_{i+r}, \ldots, k_{i-1})
\label{LoopFact} \\
&& \hskip-.6cm \null 
 + A_{r+1}^{\tree}(k_i, \ldots, k_{i+r-1}, K^h) \, {i\over K^2} \,
   A_{n-r+1}^{\tree}((-K)^{-h}, k_{i+r}, \ldots, k_{i-1}) \, 
      \cg\,  \Fact_n(K^2;k_1, \ldots, k_n) \Biggr] \,,
\nonumber
\end{eqnarray}
where the one-loop {\it factorization function} $\Fact_n$ is
independent of helicities.  (The precise form of $\Fact_n$ will not
concern us here.)  For supersymmetric NMHV amplitudes, if the number
of negative-helicity gluons in the set $\{ i, i+1, \ldots, i+r-1 \}$ 
is either 0 or 3, then the right-hand side of \eqn{LoopFact} 
vanishes.  This happens because one of the two amplitudes in 
each term then has at most one negative-helicity gluon, and such amplitudes 
are zero in supersymmetric theories.  If the number of negative-helicity 
gluons on one side of the pole is 1 or 2, then exactly one of the two values
of the intermediate helicity $h$ gives a nonvanishing contribution 
to each term in~\eqn{LoopFact}, of the form MHV~$\times$~MHV.  
The box coefficients for the MHV one-loop
amplitudes are simply given by the tree 
amplitude~(\ref{PT}) in the case that the 
box is an easy two-mass, or one-mass box, and zero 
otherwise~\cite{Neq4Oneloop}.
Hence we expect to find a limiting behavior for the NMHV box
coefficients of,
\be
A_{r+1}^{\tree}(k_i, \ldots, k_{i+r-1}, K^h)
 {1\over K^2} 
 A_{n-r+1}^{\tree}((-K)^{-h}, k_{i+r}, \ldots, k_{i-1})  \,,
\label{boxcoeffmultifact}
\ee
in appropriate nonvanishing limits.

Before addressing which limits should be nonvanishing, we inspect
the two possible sources of multi-particle poles in the building blocks
$c^{3{\rm m}}$ given in \eqn{threemassc}.
The first source is the manifest $1/K_B^2$ factor, where $B$ is
the mass diagonally opposite the massless leg $s$.  
The second source only arises when either mass $A$ or $C$ degenerates
to a single massless leg.  Suppose $A$ has a single element.
Then we can simplify one of the spinorial denominator factors
in~\eqn{threemassc} to
\be
\sandmp{s}.{\Ksl_C\Ksl_B}.{A_{-1}}
= \sandmp{s}.{(\Ksl_C+\ksl_s)(\ksl_A+\Ksl_B)}.{A}
= - (K_C+k_s)^2 \spa{s}.{A} \,,
\label{hiddenmulti}
\ee
exposing the second type of multi-particle pole.

Next we examine the residues of these poles.
First suppose all negative helicities are on one side of the pole.
In the case of the $1/K_B^2$ pole, this means that either
$m_{1,2,3} \in B$, for which $\Numer$ vanishes according to 
\eqn{NBBB}, or else no negative helicity belongs to $B$, 
for which eqs.~(\ref{NAAA}), (\ref{NAAC}) and (\ref{NsAC})
show that the would-be pole is killed by factors of $K_B^2$ in
the numerator $\Numer^4$.  
In the case of the $1/(K_C+k_s)^2$ pole
from \eqn{hiddenmulti}, when all negative helicities are on one side
the factor $\Numer$ vanishes identically, except for the case 
$m_{1,2} \in B$, $m_{3} \in A$ in \eqn{NABB}, for which it is
proportional to the vanishing denominator:
$\sandmp{s}.{\Ksl_C\Ksl_B}.{m_3} = \sandmp{s}.{\Ksl_C\Ksl_B}.{A_{-1}}$.
Thus we have verified the `trivial case' where no multi-particle
pole was expected.

If one or two negative helicities are on one side of the pole,
for either the $1/K_B^2$ pole or the $1/(K_C+k_s)^2$ pole,
then we find that such a $c^{3{\rm m}}$ coefficient always
has the limit~(\ref{boxcoeffmultifact}).  For example,
if $m_{1,2} \in B$, $m_3 \in A$, then in the limit $K_B^2 \to 0$
we have,
\be
\Numer = \spa{m_1}.{m_2} \sandmp{s}.{\Ksl_C \Ksl_B}.{m_3}
\to \spa{m_1}.{m_2} \sandmm{s}.{\Ksl_C}.{B} \spa{B}.{m_3} \,.
\ee
Because $\sandmm{s}.{\Ksl_A}.{B} = - \sandmm{s}.{\Ksl_C}.{B}$
in this limit, the four spinor strings in the denominator of
\eqn{threemassc} cancel the factor of ${\sandmm{s}.{\Ksl_C}.{B}}^4$ 
from the numerator.  Thus $c^{3{\rm m}}$ behaves as,
\bea
c^{3{\rm m}} &\to& { 1 \over \spa1.2\spa2.3\cdots\spa{n}.1 \ K_B^2 }
{ {\spa{m_1}.{m_2}}^4 \spa{B}.{m_3}^4 \spa{A_{-1}}.{B_1} \spa{B_{-1}}.{C_1} 
 \over \spa{B}.{A_{-1}} \spa{B}.{B_1}  \spa{B}.{B_{-1}} \spa{B}.{C_1} }
\nonumber \\
&=& i { {\spa{B}.{m_3}}^4 \over \spa1.2 \cdots \spa{A_{-1}}.{B} 
\spa{B}.{C_1} \cdots \spa{n}.{1} }
\times {1\over K_B^2} \times 
i { {\spa{m_1}.{m_2}}^4 \over \spa{(-B)}.{B_1} \cdots \spa{B_{-1}}.{(-B)} }
\,,
\label{NABBmulti}
\eea
which is the desired form~(\ref{boxcoeffmultifact}).
The other partitionings of negative helicities work similarly, 
for both the $1/K_B^2$ and $1/(K_C+k_s)^2$ poles.

No three-mass boxes appear in the residues of the
multi-particle poles, because the $K_B^2 \to 0$ limit forces
the three-mass boxes containing that pole to become easy two-mass boxes in
the daughter amplitude.  Indeed, this behavior directly reproduces
the daughter easy two-mass boxes where $K = K_B$ is a singlet leg.
If instead $K$ is buried in a massive leg of the daughter easy two-mass box,
then this type of term typically originates from the
$1/K_B^2$ pole in a $c^{3{\rm m}}$ contributing to an easy two-mass box
coefficient~(\ref{twomassec})
in the parent amplitude.  Figure~\ref{c2meFigure}, illustrating
the double sum for $c^{{\rm 2m}e}$, exposes the poles:   Locations of $K$
counterclockwise from $s_1$ and clockwise from $s_2$ generally come
from the left sum in \fig{c2meFigure} (in the parent amplitude),
where $K$ can be identified with $K_B$ for some set $B$.
The ones clockwise from $s_1$ and counterclockwise from $s_2$ 
generally come from the right sum.  However, if $K$ is 
counterclockwise from $s_1$ and {\it adjacent} to it
(or similarly located with respect to $s_2$), then 
$K$ cannot be identified with a $K_B$.  There is no room
for even a single-leg $A$ or $C$ argument.
In this case, the daughter term arises from 
a pole of the type $1/(K_C+k_s)^2$, in a degenerate case of
$c^{3{\rm m}}$ where $A$ or $C$ has a single element.
(In the alternative representation of easy two-mass box coefficients
with reversed `handedness', the sources of some poles get exchanged.)

Additional checks are possible from other methods of performing the 
calculation.
Very recently, Britto, Cachazo and Feng have found an
elegant and effective means for obtaining box coefficients from
the quadruple cuts, even when legs of the box integrals are massless,
by utilizing a $(--++)$ signature for space-time~\cite{BCF4m}.  
We have applied this technique to some of our coefficients,
and have found agreement with our direct calculation.


\section{New Representations of Tree Amplitudes}
\label{TreeSection}

\begin{figure}[t]
\centerline{\epsfxsize 5. truein \epsfbox{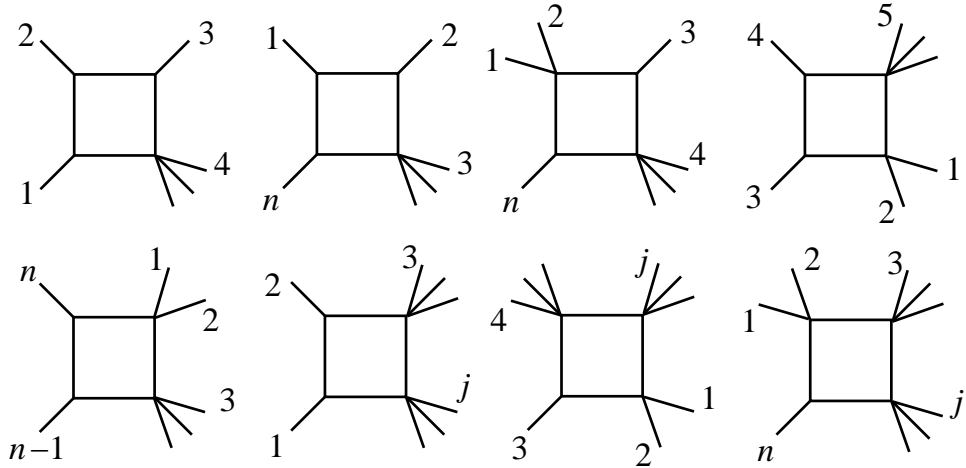}}
\caption[a]{\small The boxes whose coefficients combine to give one of the new
representations of any one-loop $n$-point tree amplitude. These boxes are
the ones with infrared singularities of the form
$\ln(-s_{12})/\e$.}
\label{TreescoeffsFigure}
\end{figure}

As we have discussed in \sect{CalculationSection}, the 
infrared consistency equation (\ref{OneLoopIRPoles}) can be
used to compute some of the box coefficients.  As we have
seen in \sect{ResultsSection}, it yields a simple and regular
form for the resulting coefficients.
As a by-product, it also
yields a variety of new representations of the $n$-point tree
amplitudes.  Because the fermions and scalars do not contribute to
$n$-gluon tree amplitudes, these representations are valid in all
massless gauge theories, including QCD. To instantiate
one of these representations,
we simply collect the coefficients of all boxes with an infrared singularity in
any given two-particle invariant.  For example, focusing on the
$\ln(-s_{12})/\eps$ singularity,
and inspecting \fig{TreescoeffsFigure},
we obtain, for any helicity configuration, the
following form for the $n$-gluon tree amplitude,
\begin{equation}
2 A_n^{\rm tree} = 
                     2 c_{1234}                             
                   + 2 c_{123n}                             
                   - 2 c_{134n}                             
                   - c_{1345}                               
                   - c_{13(n-1)n}                           
                   +  \sum_{j=5}^{n-1} c_{123j}             
                   - \sum_{j=6}^{n-1} c_{134j}              
                   - \sum_{j=5}^{n-2} c_{13jn} \,.          
\hskip .3 cm 
\label{NewTreeExample}
\end{equation}
(When an amplitude has more than three negative-helicity gluons,
four-mass boxes will
appear in the one-loop amplitude, however because these boxes
are infrared finite they do not contribute to
\eqn{NewTreeExample}.)  Other representations may be obtained
by cyclicly permuting the labels in \eqn{NewTreeExample}.  We may also
shift terms around by using the $n(n-5)/2$ additional identities
obtained from the absence of infrared singularities in multi-particle 
channels,
namely $\ln(-s_{i..j})/\eps$, $j > i+1$.

These new representations of $n$-point tree amplitudes have features
reminiscent of amplitudes built from CSW diagrams~\cite{CSW}.  In
particular, for the NMHV case most terms have only a single
multi-particle pole, coming from the $1/K_B^2$ in $c^{\rm 3m}$ in
\eqn{threemassc}.  The CSW diagrams have the same property.  The
exception is if a $c^{\rm 3m}$ appearing in the expressions for
$c^{{\rm 2m} h}$ or $c^{{\rm 2m} e}$ in \eqns{twomasshc}{twomassec}
has `degenerate' kinematics where one of the masses vanishes, then,
as mentioned in \sect{ConsistencySection},
such terms can contain two different multi-particle poles.
The appearance of spurious denominators of the form
$\sandmp{s}.{\Ksl_C\Ksl_B}.{A_{-1}}$ is again reminiscent of the
CSW approach.  (These denominators are `spurious' in the
sense that the $S$-matrix has no singularities corresponding to their
vanishing.)  However, in the CSW representation of the tree amplitudes,
the spurious denominators depend explicitly on a single arbitrary reference 
momentum $\eta$; for example, strings like 
$\spa{C^*}.{B^*} \equiv \sandpm{\eta}.{\Ksl_C\Ksl_B}.{\eta}$ 
can appear (although this particular length string first appears
in N$^2$MHV tree amplitudes). 
In the box-coefficient representation, only physical 
external momenta appear, and no single external momentum can play 
the role of $\eta$ in all terms.
The variety of different representations for the tree
amplitudes, each with its own set of spurious denominators, suggests
the existence of an even more general formalism for obtaining tree
amplitudes than the one found by CSW.


\section{Twistor-Space Properties}
\label{TwistorSection}

\def\psl{\not{\hbox{\kern-2.3pt $p$}}}
\def\tlambda{\tilde\lambda}
\def\adot{{\dot a}}
\def\bdot{{\dot b}}
\def\tspa#1.#2{\left\langle#1,#2\right\rangle}
\def\tspb#1.#2{\left[#1,#2\right]}

\subsection{Overview}
\label{TwistorOverviewSubsection}

The target space for Witten's candidate topological string theory is
$\CP^{3|4}$, otherwise called projective (super-)twistor space.  Points
in twistor space correspond to null momenta or equivalently to light cones
in space-time.  The correspondence is specified by a `half-Fourier'
transform.  More precisely, if we represent a null momentum by the
tensor product of a spinor $\lambda^a$ and a conjugate spinor 
$\tlambda^\adot$,
then twistor quantities are obtained by Fourier-transforming with respect
to all the $\tlambda^\adot$.

Amplitudes in twistor space, as it turns out, have rather simple
properties.  At tree level, they are non-vanishing only on certain curves.
This implies that they contain factors of delta functions (or derivatives
thereof) whose arguments are the characteristic equations for the curves.
The coefficients of the delta functions, however, have been quite
difficult to calculate directly from the topological string.  

As Witten pointed out in his original
paper~\cite{WittenTopologicalString}, however, we do not need the
twistor-space amplitudes in order to establish the structure of the delta
functions they contain.  In momentum space, the Fourier transform turns
the characteristic-equation
polynomials into differential operators (polynomial in the
$\lambda_i$, and derivatives with respect to the $\tlambda_i$), which will
annihilate the amplitude.  One particularly useful building block for
these differential operators is the line annihilation operator, expressing
the condition that three points in twistor space lie on a common `line' or
$\CP^1$.  If the coordinates of the three points, labeled $i,j,k$, 
are $Z^I_i = (\lambda^a_i,\mu^{\adot}_i)$, {\it etc.}, then 
the appropriate condition is
\begin{equation}
\epsilon_{IJKL} Z^I_i Z^J_j Z^K_k = 0\,,
\end{equation}
for all choices of $L$.  Choosing $L=\adot$, and translating this
equation back to momentum space using the identification 
$\mu^\adot \leftrightarrow -i\partial/\partial\tlambda_\adot$, we 
obtain the operator,
\be
F_{ijk} = \spa{i}.{j} 
{\partial\over\partial\tlambda_k}
+\spa{j}.{k} 
{\partial\over\partial\tlambda_i}
+\spa{k}.{i} 
{\partial\over\partial\tlambda_j} \,.
\label{Fdef}
\ee

Two important sufficient conditions for $F_{ijk}$ to annihilate
an expression, {\it i.e.} for it to have support only when $i,j,k$ lie on 
a line in twistor space, are~\cite{WittenTopologicalString}
\begin{enumerate}
\item The expression is completely independent of $\tlambda_i$, $\tlambda_j$,
and $\tlambda_k$, or
\item $\tlambda_i$, $\tlambda_j$, $\tlambda_k$ appear only via a sum of 
momenta containing them, of the form 
\be
P^{a\adot} = (\cdots + k_i + k_j + k_k + \cdots)^{a\adot}
  = \cdots + \lambda_i^a\tlambda_i^\adot 
           + \lambda_j^a\tlambda_j^\adot
           + \lambda_k^a\tlambda_k^\adot
           + \cdots
\label{ijkMomSum}
\ee
\end{enumerate}
The first condition is obvious from the definition~(\ref{Fdef}); 
the second holds because of the Schouten
identity,
\be
  \spa{i}.{j} \lambda_k 
+ \spa{j}.{k} \lambda_i
+ \spa{k}.{i} \lambda_j = 0\,.
\label{Schouten}
\ee

The tree-level MHV amplitude~(\ref{PT}), for example, 
is annihilated by $F_{ijk}$,
because it is independent of the $\tlambda_i$.  Any possible delta
functions vanish for generic momenta, because they take the form
$\delta(\spa{i}.{j})$.  At one loop, 
Cachazo, Svr\v{c}ek, and Witten~\cite{CSWIII} pointed out that
such delta
functions, arising from the spinor analog of the fact that 
$\partial_{\overline z}\, (1/z)\neq 0$, do arise.  They must be taken
into account for a proper analysis of the twistor-space structure
of amplitudes.

We will not compute the relevant `holomorphic anomaly' terms for the amplitudes
in this paper, and so we will not be able to fully exhibit their
twistor-space structure.  While the `anomaly' terms enter into the
action of the differential operators on the box integrals, their
action on the coefficients is unaffected by it.  The properties
of the coefficients are also important, so we focus on these.

In addition to the line operator $F_{ijk}$, we will employ the 
planar operator~\cite{WittenTopologicalString},
\be
K_{ijkl} \equiv \epsilon_{IJKL} Z^I_i Z^J_j Z^K_k Z^L_l
 = \spa{i}.{j} \varepsilon^{\adot\bdot}
   {\partial\over\partial\tlambda_k^\adot}
   {\partial\over\partial\tlambda_l^\bdot}
  \pm \hbox{[5 permutations]} \,,
\label{Kdef}
\ee
whose vanishing implies that four points lie in a plane 
(or $\CP^2$) in twistor space.


\subsection{Twistor properties of three-mass box coefficients}
\label{ThreeMassSubsection}

As discussed in \sect{ResultsSection}, the three-mass box coefficients
$c^{3{\rm m}}$ given in \eqn{threemassc}
are the basic building blocks for the NMHV amplitudes.
All other box coefficients can be expressed as sums of various $c^{3{\rm m}}$.
Therefore we need only determine the
twistor-space properties of the three-mass box coefficients, in order 
to obtain the general twistor-space properties of all the box
coefficients.

The most general twistor-space property of the NMHV box coefficients
is that {\it all points lie in a plane}.  That is, $K_{mnpq}$ for
every choice of $m,n,p,q$ annihilates every one-term coefficient, and
hence, by linearity, it annihilates every box coefficient $c_{ijkl}$.
We first observed the planarity of a special class of three-mass box
coefficients in ref.~\cite{Neq47}.  The complete co-planarity for all
coefficients was proven for general one-loop NMHV
amplitudes~\cite{Neq47,BCFcoplanar}, along the same lines used by
Cachazo to demonstrate a certain degree of
co-linearity~\cite{Cachazo}.

Since we have computed all NMHV coefficients, it is straightforward to
confirm directly that the required planarity property holds.  The
co-planarity of $s$, $A$, and $C$ can be demonstrated relatively
easily, however the co-planarity with $B$ requires more work.  In
\app{CoPlanDemoSection} we present an analytic demonstration of
planarity of all three-mass box coefficients.  This in turn 
implies that all remaining coefficients are sums of planar functions
since 
they are sums of three-mass coefficients and their degenerate limits.

\begin{figure}[t]
\centerline{\epsfxsize 5.5 truein \epsfbox{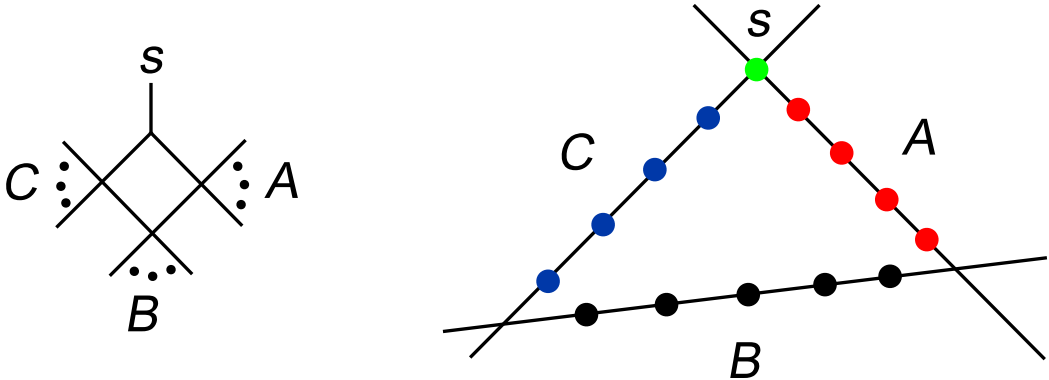}}
\caption[a]{\small The twistor-space configuration for
the three-mass-box coefficients $c^{3{\rm m}}$ given in 
\eqn{threemassc} is depicted on the right.  The corresponding
three-mass integral is shown on the left.  All points lie in a plane.}
\label{allntwistFigure}
\end{figure}

Another intriguing property of the non-vanishing coefficients is the
universality of the distribution of points on the three lines,
independent of the identity of the three negative-helicity legs.  As
illustrated in \fig{allntwistFigure}, the location of the points is
entirely dictated by the three-mass box under consideration.  The
helicity independence may be understood simply by considering the
triple cuts of the three-mass box. As argued by Cachazo~\cite{Cachazo}
for the more standard double cut, the fact that the holomorphic
anomaly freezes the phase-space integral~\cite{CSWIII,BBKR} implies
that the box integral coefficients are annihilated by the same
co-linear operators $F_{ijk}$ that annihilate the trees on either side
of the cut.  A similar argument for the triple cut shows that the
coefficient of a three-mass box must be annihilated by the same
co-linear operators that annihilate each of the three trees.  For all
non-vanishing three-mass box coefficients the three tree amplitudes
appearing in the triple cut are all MHV.  Thus for each of the three
clusters $A$, $B$ and $C$ the points must be on a line, independent of
the location of the negative helicities, since this is a property of
the MHV tree amplitudes. (The presence of the point $s$ at the
intersection of the two lines containing $A$ and $C$ follows from the
existence of two distinct triple cuts: one cut where $s$ is a point in
the tree amplitude containing the $A$ and one where it is a point in
the tree containing the $C$.)  We find it extremely appealing that the
simplicity of the structure displayed in \fig{allntwistFigure}
is reflected in the NMHV one-loop amplitudes computed here.

\begin{figure}[t]
\centerline{\epsfxsize 5.5 truein \epsfbox{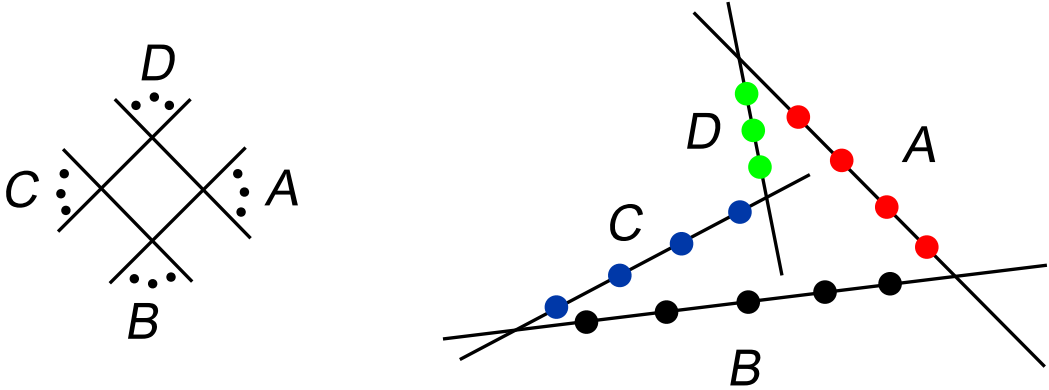}}
\caption[a]{\small The twistor-space configuration for the
four-mass-box coefficient of an N$^2$MHV amplitude is depicted on the
right.  The corresponding four-mass integral is shown on the left.
The points need not lie in a plane, but lie in the intersection of two
planes.}
\label{nnmhvtwistFigure}
\end{figure}

It is worth noting that a similar property holds for the N$^2$MHV
(next-to-next-to-MHV)
$n$-point amplitudes. The quadruple cut shows that each cluster in a
four-mass box coefficient must always lie on a straight line, since
once again each of the four clusters in the quadruple cut is an MHV
tree amplitude if the coefficient does not vanish.  Moreover, in the
triple cut either MHV trees or NMHV trees made up of 
nearest-neighbor clusters are found. The NMHV trees formed by two 
nearest-neighbor clusters are supported on two intersecting lines.
Stepping through the four triple cuts then implies that 
nearest-neighbor clusters are localized on intersecting lines.  This
picture agrees with the properties of the eight-point coefficient of
the four-mass box obtained by Britto, Cachazo and Feng~\cite{BCF4m}.
For larger numbers of negative-helicity legs, one can no longer
conclude that the points in each cluster lie on straight lines, because
the quadruple cuts are no longer products of MHV amplitudes.  We may
expect the structure of \fig{nnmhvtwistFigure} to generalize, however,
with each of the line segments replaced by the twistor-space duals to 
higher-degree vertices~\cite{Gukov,BBK}, that is appropriate collections
of intersecting line segments.

The twistor-space structure of the one- and two-mass box coefficients
in the NMHV amplitudes is of course completely determined by the
structure of the three-mass coefficients, using 
eqns.~(\ref{twomasshc}), (\ref{twomassec}),
and~(\ref{onemassc}).  Each term in these sums will have its support
in a plane (that is, a $\CP^2$) in twistor space, though not
necessarily the same plane for all terms.  Each term has all points
lying on three lines within a single plane; and one of the
intersections of the lines always contains one of the $n$ points.


\section{Conclusions and Outlook}
\label{ConclusionSection}

In this paper we computed all the next-to-MHV one-loop gluon
amplitudes in $\NeqFour$ super-Yang--Mills theory.
The coefficients of the box integral functions appearing
in the amplitudes can be written as simple terms built out of spinor
strings, or sums of such terms, where  each term exhibits a simple
twistor-space structure.

To obtain the four-, three- and hard two-mass box coefficients we used
generalized cuts~\cite{ZFourPartons,TwoLoopUnitarity,TwoLoopSplitting,Neq47}
in the unitarity
method~\cite{Neq4Oneloop,Neq1Oneloop,UnitarityMachinery,TwoLoopSplitting}.
The four-mass box coefficients all vanish in the NMHV case, as easily
determined from quadruple cuts.  We also showed how to obtain hard
two-mass box coefficients from the three-mass box coefficients using
the known behavior of amplitudes as external momenta become soft.  The
easy two-mass and one-mass coefficients were then obtained efficiently
by solving the constraints that the infrared singularities of the
amplitudes (as regulated by dimensional regularization) match the
known universal form~\cite{UniversalIR}. We
also confirmed some of these coefficients
by direct computation of ordinary cuts.  For odd
$n$, the infrared consistency equations suffice to obtain all these
coefficients.  For even $n$, a lone infrared consistency equation is
missing.  We computed the missing coefficient from the requirement
that amplitudes have the correct collinear limits.  The solution to
the infrared consistency equations yields a very regular form for the
easy two-mass and one-mass coefficients. 

We may also apply the structure of the generalized cuts to determine
some terms of the corresponding amplitudes in QCD.
Many types of integral functions beyond those appearing
in the $\NeqFour$ results contribute
to the full QCD results --- scalar triangle integrals, scalar and 
tensor bubble integrals --- and their coefficients are of course
undetermined by the calculations in this paper.  However, many
of the box coefficients are the same as those given 
in~\sect{ResultsSection}.  As a particular
example, the coefficient of
any box integral where only positive-helicity gluons form
one of the massive legs is identical in QCD 
and the $\NeqFour$ theory (and in the pure-glue theory as well).

The infrared consistency equations also provide us with new
representations of $n$-gluon NMHV tree-level amplitudes.  The form in
which the amplitudes appear is similar to the one obtained using MHV
vertices~\cite{CSW}.  There are, however, a number of differences. The
variety of different representations for the tree amplitudes, each
with their own set of spurious denominators, suggests a more general
formalism for obtaining tree amplitudes exists than the one found by
Cachazo, Svr\v{c}ek, and Witten.

We also performed a variety of checks, including verifying that 
amplitudes have the correct behavior in various
collinear, soft, and multi-particle factorization limits.
 As a final check, we also used the observation
of Britto, Cachazo, and Feng~\cite{BCF4m} last week,
that generalized quadruple cuts
freeze the loop integrals, allowing for an elegant and simple
algebraic solution of box coefficients.  The utilization of
a non-conventional $(--++)$-signature metric, 
allows quadruple cuts to be applied even when some
of the box function's external legs are massless.

The planarity of the NMHV box coefficients is a very intriguing
result.  The complete planarity was demonstrated for general one-loop
NMHV amplitudes~\cite{Neq47,BCFcoplanar}, along the same lines used by
Cachazo to demonstrate a certain degree of
co-linearity~\cite{Cachazo}.  The explicit calculation of the
coefficients presented here confirms these arguments.  In twistor
space, the points in the three-mass box coefficients fall into three
lines lying in a plane, and two of the lines always intersect at one
of the $n$ points, as depicted in~\fig{allntwistFigure}.  In
particular, the split-up is independent of the particular NMHV
helicity configuration, and only depends on the kinematics of the
particular three-mass box.  As described in the paper, this structure
is easy to understand using the generalized cuts together with the
twistor-space properties of the tree amplitudes appearing in the cuts.

The simplicity of the amplitudes, found here and in
refs.~\cite{Neq4Oneloop,Neq1Oneloop,Cachazo,BCF7,Neq47,DunbarNeq1,
BCF4m,BBSTNeq0}, suggests that the complete one-loop $S$-matrix of all
four-dimensional cut-constructible gauge theories will be obtained soon.
Their simple twistor-space structure also suggests the search
for a string interpretation will be fruitful.


\section*{Acknowledgments}

We thank Radu Roiban, Marcus Spradlin and Anastasia Volovich for
helpful discussions on the twistor-space structure of four-mass box
coefficients. We also thank Iosif Bena for helpful discussions on
general twistor-space properties. We also thank Academic Technology
Services at UCLA for computer support. Some of the diagrams in this
paper were constructed with {\tt JaxoDraw}~\cite{Jaxodraw}.  The {\it
Service de Physique Th\'eorique\/} is a laboratory of the {\it
Direction des Sciences de la Mati\`ere\/} of the {\it Commissariat \`a
l'Energie Atomique\/} of France.

\appendix

\section{Explicit Demonstration of Planarity of Coefficients}
\label{CoPlanDemoSection}

The planarity of any box coefficient in any one-loop NMHV
amplitude has already been proven on general
grounds~\cite{Neq47,BCFcoplanar}.  Nevertheless, it is interesting to
see how it works explicitly, now that the complete NMHV results
are known.  Because every NMHV box coefficient is a sum of the
three-mass box coefficients $c^{3{\rm m}}$ given in \eqn{threemassc}
(and the co-planarity operator $K$ is a linear operator), it suffices
to show that this expression is completely planar; that is, it has its
support in a $\CP^2$ subspace.

First we recall the two important sufficient conditions for $F_{ijk}$ 
to annihilate an expression, described below \eqn{Fdef}.
Using these, the only way dependence on anti-holomorphic spinors $\tlambda_i$
appear in \eqn{threemassc} is via $\Ksl_A$, $\Ksl_B$, and $\Ksl_C$.
Furthermore, $\Ksl_A$ and $\Ksl_C$ always appear next to $\langle s^- |$,
so that they may be re-written as $\ksl_s+\Ksl_A$ and $\ksl_s+\Ksl_C$,
respectively.  Thus, using \eqn{ijkMomSum}, we see that $c^{3{\rm m}}$ in 
\eqn{threemassc} has support only when all points in each of the following
three sets are co-linear: 
$\{ s \} \cup A$; $B$; and $\{ s \} \cup C$.

The co-linear constraints are shown in \fig{allntwistFigure}.
The point $s$ belongs to two lines, $A$ and $C$.
This fact implies that lines $A$ and $C$ lie in a plane.
Hence our task is to show that line $B$ also lies in this plane.
It suffices to show that 
\be
K_{a_1a_2b_1b_2} c^{3{\rm m}} = 0 \,,
\label{Konc}
\ee
for any two points $a_1,a_2 \in A$ and any two points $b_1,b_2 \in B$.
We can use momentum conservation to replace $K_A \to - k_s - K_B - K_C$
in \eqn{threemassc}.  Then the terms in $K_{a_1a_2b_1b_2}$ containing
derivatives with respect to $\tlambda_{a_1}$ and $\tlambda_{a_2}$
vanish, and \eqn{Konc} reduces to 
\be
\spa{a_1}.{a_2} \pol^{\aldot\bedot} 
    { \del \over \del \tlambda_{b_1}^\aldot }
    { \del \over \del \tlambda_{b_2}^\bedot } c^{3{\rm m}} = 0 \,.
\label{Koncnew}
\ee
So we just need to show that the double derivative in \eqn{Koncnew}
vanishes.

The first derivative is simple to evaluate, using
\bea 
{ \del \over \del \tlambda_{b_i}^\aldot } 
 \sandmp{s}.{\Ksl_C\Ksl_B}.{X}
  &=&   \spa{b_i}.{X} ( \langle s^- | \Ksl_C )_{\aldot}  \,,
\label{sCBXderiv} \\
{ \del \over \del \tlambda_{b_i}^\aldot } 
 \sandmp{s}.{\Ksl_A\Ksl_B}.{X}
  &=&   - \spa{b_i}.{X} ( \langle s^- | \Ksl_C )_{\aldot}
        - \spa{s}.{X} ( \langle b_i^- | \Ksl_B )_{\aldot} \,,
\label{sABXderiv} \\
{ \del \over \del \tlambda_{b_i}^\aldot } K_B^2
  &=& ( \langle b_i^- | \Ksl_B )_{\aldot} \,.
\label{kBsqderiv}
\eea
The derivative depends on the helicity configuration, through $\Numer$.
Here we will present the most complicated case, $m_1 \in A$, $m_2 \in B$, 
$m_3 \in C$, for which $\Numer$ is given by \eqn{NABC}.
The other cases can be worked out analogously.
We find that
\be
{ \del \over \del \tlambda_{b_1}^\aldot } c^{3{\rm m}}
= V_{\aldot}(b_1)\times c^{3{\rm m}}  \,,
\label{deriv1}
\ee
where
\bea
V_{\aldot}(b_1) &=& { 4 \over \Numer } \Bigl[
 ( \spa{m_3}.{m_2} \spa{b_1}.{m_1}
 - \spa{m_1}.{m_2} \spa{b_1}.{m_3} )  ( \langle s^- | \Ksl_C )_{\aldot} 
- \spa{m_1}.{m_2} \spa{s}.{m_3}  ( \langle b_1^- | \Ksl_B )_{\aldot} \Bigr]
\nonumber \\
&&\hskip0.0cm
- { ( \langle b_1^- | \Ksl_B )_{\aldot} \over K_B^2 }
- { \spa{b_1}.{A_{-1}} ( \langle s^- | \Ksl_C )_{\aldot}
    \over \sandmp{s}.{\Ksl_C\Ksl_B}.{A_{-1}} }
- { \spa{b_1}.{B_{1}} ( \langle s^- | \Ksl_C )_{\aldot}
    \over \sandmp{s}.{\Ksl_C\Ksl_B}.{B_{1}} }
\nonumber \\
&&\hskip0.0cm
+ { \spa{b_1}.{B_{-1}} ( \langle s^- | \Ksl_C )_{\aldot}
  + \spa{s}.{B_{-1}} ( \langle b_1^- | \Ksl_B )_{\aldot}
    \over \sandmp{s}.{\Ksl_A\Ksl_B}.{B_{-1}} }
\nonumber \\
&&\hskip0.0cm
+ { \spa{b_1}.{C_{1}} ( \langle s^- | \Ksl_C )_{\aldot}
  + \spa{s}.{C_{1}} ( \langle b_1^- | \Ksl_B )_{\aldot}
    \over \sandmp{s}.{\Ksl_A\Ksl_B}.{C_{1}} } \,.
\label{Vb1def}
\eea

The second derivative has two types of terms,
\be
\pol^{\aldot\bedot} 
    { \del \over \del \tlambda_{b_1}^\aldot }
    { \del \over \del \tlambda_{b_2}^\bedot } c^{3{\rm m}}
= \Biggl[ \pol^{\aldot\bedot} 
  { \del \over \del \tlambda_{b_2}^\bedot } V_{\aldot}(b_1)
   + \pol^{\aldot\bedot} V_{\aldot}(b_1) V_{\bedot}(b_2) \Biggr]
   \times c^{3{\rm m}}  \,,
\label{deriv2types}
\ee
Note that
\be
\pol^{\aldot\bedot}  ( \langle s^- | \Ksl_C )_{\aldot}
                     ( \langle s^- | \Ksl_C )_{\bedot}
= - \sandmp{s}.{\Ksl_C\Ksl_C}.{s} 
= - K_C^2 \spa{s}.{s} = 0.
\label{CCvanish}
\ee
Using this fact, it is easy to see that in the first type of terms ---
those coming from the derivative of $V_{\aldot}(b_1)$ --- 
the terms containing $\sandmp{s}.{\Ksl_C\Ksl_B}.{X}$
do not contribute.  The term containing $K_B^2$ gives
\bea
 \pol^{\aldot\bedot} 
  { \del \over \del \tlambda_{b_2}^\bedot }  \biggl[
- { ( \langle b_1^- | \Ksl_B )_{\aldot} \over K_B^2 } \biggr]
&=& \pol^{\aldot\bedot} \Biggl[ 
{1 \over (K_B^2)^2 } ( \langle b_1^- | \Ksl_B )_{\aldot}
                     ( \langle b_2^- | \Ksl_B )_{\bedot}
- {1 \over K_B^2 } (-1) \spa{b_1}.{b_2} \pol_{\bedot\aldot} \Biggr]
\nonumber \\
&=& { \spa{b_1}.{b_2} \over K_B^2 } \,.
\label{kBsqtype1}
\eea
A slightly more complicated term is
\bea
 &&
\pol^{\aldot\bedot} 
  { \del \over \del \tlambda_{b_2}^\bedot } 
 { \spa{b_1}.{B_{-1}} ( \langle s^- | \Ksl_C )_{\aldot}
  + \spa{s}.{B_{-1}} ( \langle b_1^- | \Ksl_B )_{\aldot}
    \over \sandmp{s}.{\Ksl_A\Ksl_B}.{B_{-1}} }
\nonumber \\
&=& 
\pol^{\aldot\bedot} 
\Biggl\{  { 1 \over {\sandmp{s}.{\Ksl_A\Ksl_B}.{B_{-1}}}^2 }
   \Bigl[ \spa{b_1}.{B_{-1}} ( \langle s^- | \Ksl_C )_{\aldot}
  + \spa{s}.{B_{-1}} ( \langle b_1^- | \Ksl_B )_{\aldot} \Bigr]
\nonumber \\
&& 
  \hphantom{
     \pol^{\aldot\bedot} 
     \Biggl\{  { 1 \over {\sandmp{s}.{\Ksl_A\Ksl_B}.{B_{-1}}}^2 } }
\times 
\Bigl[ \spa{b_2}.{B_{-1}} ( \langle s^- | \Ksl_C )_{\bedot}
  + \spa{s}.{B_{-1}} ( \langle b_2^- | \Ksl_B )_{\bedot} \Bigr]
\nonumber \\
&& \hskip0.9cm 
-  { \pol_{\bedot\aldot} \spa{b_1}.{b_2} \spa{s}.{B_{-1}}
   \over \sandmp{s}.{\Ksl_A\Ksl_B}.{B_{-1}} } 
  \Biggr\}
\nonumber \\
&=& 
 - { \spa{s}.{B_{-1}} \over {\sandmp{s}.{\Ksl_A\Ksl_B}.{B_{-1}}}^2 }
 \Bigl[ \spa{b_1}.{B_{-1}} \sandmp{s}.{\Ksl_C\Ksl_B}.{b_2} 
      - \spa{b_2}.{B_{-1}} \sandmp{s}.{\Ksl_C\Ksl_B}.{b_1} 
\nonumber \\
&& 
 \hphantom{
 - { \spa{s}.{B_{-1}} \over {\sandmp{s}.{\Ksl_A\Ksl_B}.{B_{-1}}}^2 } \Bigl[]}
      + \sandmp{b_1}.{\Ksl_B\Ksl_B}.{b_2}  \spa{s}.{B_{-1}} \Bigr]
\nonumber \\
&& \hskip0.0cm 
- 2 { \spa{b_1}.{b_2} \spa{s}.{B_{-1}}
   \over \sandmp{s}.{\Ksl_A\Ksl_B}.{B_{-1}} } \,.
\label{sABX} 
\eea 
In the quantity in brackets ($[~]$) on the right-hand side 
of~\eqn{sABX}, we use the Schouten identity to combine the 
first two terms, and rewrite the third term as 
$\spa{b_1}.{b_2} \sandmp{s}.{\Ksl_B\Ksl_B}.{B_{-1}}$.
Then this quantity becomes
\be
 \spa{b_1}.{b_2} \sandmp{s}.{(\Ksl_C+\Ksl_B)\Ksl_B}.{B_{-1}} 
 = - \spa{b_1}.{b_2} \sandmp{s}.{\Ksl_A\Ksl_B}.{B_{-1}} \,.
\label{sABXbrackets}
\ee
Inserting this expression into \eqn{sABX}, we obtain,
\be
\pol^{\aldot\bedot} 
  { \del \over \del \tlambda_{b_2}^\bedot } 
 { \spa{b_1}.{B_{-1}} ( \langle s^- | \Ksl_C )_{\aldot}
  + \spa{s}.{B_{-1}} ( \langle b_1^- | \Ksl_B )_{\aldot}
    \over \sandmp{s}.{\Ksl_A\Ksl_B}.{B_{-1}} }
= - { \spa{b_1}.{b_2} \spa{s}.{B_{-1}}
   \over \sandmp{s}.{\Ksl_A\Ksl_B}.{B_{-1}} } \,.
\label{sABXfinal} 
\ee

The remaining two nontrivial terms in the derivative of 
$V_{\aldot}(b_1)$ work very similarly.  Assembling all four terms, 
we have
\be
\pol^{\aldot\bedot} 
  { \del \over \del \tlambda_{b_2}^\bedot } V_{\aldot}(b_1)
= \spa{b_1}.{b_2} \Biggl[ 
  4 { \spa{m_1}.{m_2} \spa{s}.{m_3}
      \over \sandmp{s}.{\Ksl_A\Ksl_B}.{m_3} } 
  + { 1\over K_B^2 }
  - { \spa{s}.{B_{-1}} \over \sandmp{s}.{\Ksl_A\Ksl_B}.{B_{-1}} } 
  - { \spa{s}.{C_{1}} \over \sandmp{s}.{\Ksl_A\Ksl_B}.{C_{1}} } 
 \Biggr] \,.
\label{type1final}
\ee

Terms of the second type arise from the contraction 
$\pol^{\aldot\bedot} V_{\aldot}(b_1) V_{\bedot}(b_2)$.
There are $5 \times 5 = 25$ terms, although the terms
containing two $\sandmp{s}.{\Ksl_C\Ksl_B}.{X}$ strings do not contribute.
In each of the nonvanishing terms, the Schouten identity can again 
be used to extract a factor of $\spa{b_1}.{b_2}$, and the remainder
becomes a sum of two (or sometimes just one) of the terms 
in~\eqn{type1final}.  For example, using algebra similar to that 
in \eqn{sABX}, we get,
\bea
&&
\pol^{\aldot\bedot}
\Biggl[
{ \spa{b_1}.{B_{-1}} ( \langle s^- | \Ksl_C )_{\aldot}
  + \spa{s}.{B_{-1}} ( \langle b_1^- | \Ksl_B )_{\aldot}
    \over \sandmp{s}.{\Ksl_A\Ksl_B}.{B_{-1}} }
{ \spa{b_2}.{C_{1}} ( \langle s^- | \Ksl_C )_{\bedot}
  + \spa{s}.{C_{1}} ( \langle b_2^- | \Ksl_B )_{\bedot}
    \over \sandmp{s}.{\Ksl_A\Ksl_B}.{C_{1}} }
\nonumber \\ && \hskip1.0cm
 + ( B_{-1} \lr C_{1} ) \Biggr]
\nonumber \\
&&  \hskip 2 cm = \spa{b_1}.{b_2} \Biggl[ 
    { \spa{s}.{B_{-1}} \over \sandmp{s}.{\Ksl_A\Ksl_B}.{B_{-1}} } 
  + { \spa{s}.{C_{1}} \over \sandmp{s}.{\Ksl_A\Ksl_B}.{C_{1}} } \Biggr] \,.
\label{samplexterm}
\eea
Computing and assembling all the 
$\pol^{\aldot\bedot} V_{\aldot}(b_1) V_{\bedot}(b_2)$
contributions, we get,
\bea
\pol^{\aldot\bedot} V_{\aldot}(b_1) V_{\bedot}(b_2) &=&
\spa{b_1}.{b_2} \Biggl\{ 
 { \spa{m_1}.{m_2} \spa{s}.{m_3}
      \over \sandmp{s}.{\Ksl_A\Ksl_B}.{m_3} }
\times \Bigl[ 16 - 4 - 4 - 4 - 4 - 4 \Bigr]
\nonumber \\
&& \hskip0.5cm \null
+ {1 \over K_B^2 } \times \Bigl[ 4 - 1 - 1 - 1 - 1 - 1 \Bigr]
\nonumber \\
&& \hskip0.5cm \null
+ { \spa{s}.{B_{-1}} \over \sandmp{s}.{\Ksl_A\Ksl_B}.{B_{-1}} } 
  \times \Bigl[ - 4 + 1 + 1 + 1 + 1 + 1 \Bigr]
\nonumber \\
&& \hskip0.5cm \null
+ { \spa{s}.{C_{1}} \over \sandmp{s}.{\Ksl_A\Ksl_B}.{C_{1}} } 
  \times \Bigl[ - 4 + 1 + 1 + 1 + 1 + 1 \Bigr] \Biggr\} \,,
\label{type2final}
\eea
where the 6 numbers on each line correspond to the contribution from the cross
term of the term shown with each of the 6 terms in $V_\aldot$ in 
\eqn{Vb1def}.  The sum of \eqns{type1final}{type2final} is
zero, which demonstrates, via \eqn{deriv2types}, the planarity 
of $c^{3{\rm m}}$. 


\section{Box Integrals}
\label{IntegralsAppendix}

In this appendix we collect the dimensionally-regulated integral
functions appearing in the $\NeqFour$ amplitudes; the first of 
these integral
functions was obtained from ref.~\cite{FourMassBox} and the remaining 
ones from ref.~\cite{Integrals5}.
The reader is referred to these papers for further details.
Through $\Ord(\eps^0)$, in the Euclidean region the integral
functions are
\def\hs{\hskip 2.5 cm \null}
\begin{eqnarray}
&& F^{4{\rm m}}  (s, t, K_1^2, K_2^2, K_3^2, K_4^2)
= {1\over 2}
\Biggl\{ - \Li_2\left(\hf(1-\lambda_1+\lambda_2+\rho)\right)
   + \Li_2\left(\hf(1-\lambda_1+\lambda_2-\rho)\right) \nonumber \\
 && \hs
  - \Li_2\left(
   \textstyle-{1\over2\lambda_1}(1-\lambda_1-\lambda_2-\rho)\right)
  + \Li_2\left(\textstyle-{1\over2\lambda_1}(1-\lambda_1-
    \lambda_2+\rho)\right) \nonumber \\
 && \hs - {1\over2}\ln\left({\lambda_1\over\lambda_2^2}\right)
   \ln\left({ 1+\lambda_1-\lambda_2+\rho \over 1+\lambda_1
        -\lambda_2-\rho }\right) \Biggr\} \,, 
  \label{Fboxes4m} \\
&&  F^{3{\rm m}}(s,t, K_2^2, K_3^2, K_4^2)
=  -{1\over2\e^2} \Bigl[ (-s)^{-\e} + (-t)^{-\e}
     - (-K_2^2)^{-\e} - (-K_4^2)^{-\e} \Bigr]\nonumber \\
&& \hs    - {1\over2} \ln\left({-K_2^2\over -t}\right)
                \ln\left({-K_3^2\over -t}\right)
          - {1\over2} \ln\left({-K_3^2\over -s}\right)
                \ln\left({-K_4^2\over -s}\right)  \nonumber\\
 &&\hs + \Li_2\left(1-{K_2^2\over s}\right)
   + \Li_2\left(1-{K_4^2\over t}\right) 
   -  \Li_2\left(1-{K_2^2K_4^2\over st}\right) \nonumber\\
 &&\hs + {1\over 2} \ln^2\left({-s\over -t}\right) , 
  \label{Fboxes3m} \\
&&  F^{2{\rm m} \, h} (s, t, K_3^2, K_4^2) 
=  -{1\over 2\e^2} \Bigl[ (-s)^{-\e} + 2 (-t)^{-\e}
              - (-K_3^2)^{-\e} - (-K_4^2)^{-\e} \Bigr]
\nonumber\\
&&\hs - {1\over2} \ln\left({-K_3^2\over -s}\right)
                \ln\left({-K_4^2\over -s}\right)
   + \Li_2\left(1-{K_3^2\over t}\right)
   + \Li_2\left(1-{K_4^2\over t}\right)  \nonumber\\
 &&\hs + {1\over 2} \ln^2\left({-s\over -t}\right) ,
 \label{Fboxes2mh} \\
&&  F^{2{\rm m}\, e}(s, t, K_2^2, K_4^2) 
=   -{1\over\e^2} \Bigl[ (-s)^{-\e} + (-t)^{-\e}
              - (-K_2^2)^{-\e} - (-K_4^2)^{-\e} \Bigr] \nonumber\\
 &&\hs  + \Li_2\left(1-{K_2^2\over s}\right)
    + \Li_2\left(1-{K_2^2\over t}\right)
    + \Li_2\left(1-{K_4^2\over s}\right)
    + \Li_2\left(1-{K_4^2\over t}\right) \nonumber\\
 &&\hs  - \Li_2\left(1-{K_2^2K_4^2\over st}\right)
    + {1\over2} \ln^2\left({-s\over -t}\right) ,
   \label{Fboxes2me} \\
&&  F^{1{\rm m}} (s,t, K_4^2) 
=  -{1\over\e^2} \Bigl[ (-s)^{-\e} + (-t)^{-\e} - (-K_4^2)^{-\e} \Bigr]
            \nonumber\\
&& \hs  + \Li_2\left(1-{K_4^2\over s}\right)
   + \Li_2\left(1-{K_4^2\over t}\right)
   + {1\over 2} \ln^2\left({-s\over -t}\right) + {\pi^2\over6} \, , 
  \label{Fboxes1m} \\
&&  F^{0{\rm m}}(s,t)  = 
 - {1\over\e^2} \Bigl[ (-s)^{-\e} + (-t)^{-\e} \Bigr]
  + {1\over 2} \ln^2\left({-s\over -t}\right) + {\pi^2\over 2} \,,
  \label{Fboxes0m} 
\end{eqnarray}
where the $k_i$ denote massless  momenta and the $K_i$ massive momenta.
The external momentum arguments $K_1,\ldots,K_4$ are sums
of external momenta $k_i$ from the $n$-point amplitude.  
The kinematic variables appearing in the integrals are 
\begin{equation}
s = (k_1 + k_2)^2 \, , \hs t = (k_2 + k_3)^2\,,
\end{equation}
or with $k$ relabeled as $K$ for off-shell (massive) legs.
The functions appearing in $F_4^{4 \rm m}$ are
\begin{equation}
 \rho\ \equiv\ \sqrt{1 - 2\lambda_1 - 2\lambda_2
+ \lambda_1^2 - 2\lambda_1\lambda_2 + \lambda_2^2}\ ,
\label{rdefinition}
\end{equation}
and
\begin{equation}
\lambda_1 = {K_1^2 \, K_3^2 
\over (K_1 + K_2)^2 \, (K_2 + K_3)^2 } \; , \hskip 1.5 cm
\lambda_2 = {K_2^2  \, K_4^2 \over
 (K_1 + K_2)^2 \, (K_2 + K_3)^2  } \; .
\end{equation}
We have rearranged the expressions for $F^{3{\rm m}}$
and $F^{2{\rm m} \, h}$ to make the poles in $\e$ 
more transparent.  We have also corrected some signs
in $F^{4{\rm m}}$ in ref.~\cite{Neq4Oneloop} and in the
published version of ref.~\cite{Integrals5}.


\end{document}